# Compositional and Oxygen-Vacancy Effects on Phase Stability and Electronic Properties in Ceria-Based Lanthanide High-Entropy Oxides

Mary Kathleen Caucci[a], Billy E. Yang[b], Gerald R. Bejger[c], Jacob T. Sivak[a], Christina M. Rost[c], Saeed S.I. Almishal[b], Jon-Paul Maria[b], Susan B. Sinnott [a,b,d,e,*]

[a] *Department of Chemistry, The Pennsylvania State University, University Park, PA 16802 USA*
[b] *Department of Materials Science and Engineering, The Pennsylvania State University, University Park, PA 16802 USA*
[c] *Department of Materials Science and Engineering, Virginia Polytechnic Institute and State University, Blacksburg, VA 24060 USA*
[d] *Materials Research Institute, The Pennsylvania State University, University Park, PA, 16802 USA*
[e] *Institute for Computational and Data Science, The Pennsylvania State University, University Park, PA, 16802 USA*

(Dated: December 28, 2025)

**Abstract**
Cerium-based lanthanide high-entropy oxides (LN-HEOs) are promising candidates for solid-state electrolyte (mass transport) applications due to their ability to accommodate high concentrations of oxygen vacancies while retaining a fluorite-derived structure. However, synthesis often yields undesired ordered oxygen-deficient phases, such as bixbyite, depending on composition and processing conditions. We utilize first-principles density functional theory (DFT) calculations to systematically investigate phase stability in the model system $Ce_x(YLaPrSm)_{1-x}O_{2-\delta}$, with the aim of elucidating the thermodynamic factors governing fluorite–bixbyite competition and identifying structure-property relationships to oxygen transport. By independently varying cerium concentration and oxygen vacancy content, we predict that the transition from disordered fluorite to ordered bixbyite is driven primarily by compositional and vacancy-ordering effects, rather than through changes in cation valence. Free-energy analysis reveals that at high vacancy concentrations, bixbyite is enthalpically favored due to ordered oxygen vacancies, while fluorite is stabilized at lower vacancy concentrations and higher cerium content through configurational entropy of the anion sublattice. These DFT results clarify the competing energetic contributions that control phase stability and structure–valence relationships in LN-HEOs and establishes a mechanistic framework for designing vacancy-tolerant oxide electrolytes with tunable phase behavior.

**Keywords**
Density Functional Theory; High-Entropy oxides; Lanthanide oxides; Electronic Structure; First-principles calculations; Thermodynamics

*Corresponding author

# 1. INTRODUCTION

Valence-change oxides that operate through the migration and redistribution of oxygen vacancies ($V_O$) form the foundation for novel, energy-efficient technologies, including solid-state ionics, resistive switching devices, and redox-active catalysts, by promising control over resister states through local structural and electronic rearrangements [1–10]. In these materials, local changes in oxygen coordination induce coupled structural and electronic rearrangements that directly control transport, conductivity, and functional response. Despite extensive study [1–12], a detailed understanding of how structural disorder, cation chemistry, and oxygen vacancy configuration govern ionic mobility and redox activity remain poorly understood, particularly in complex oxide systems where multiple competing interactions coexist.

High-entropy oxides (HEOs) represent a unique platform for designing oxide materials with tunable functionality for addressing these challenges. Following the introduction of entropy-stabilized oxides by Rost *et al.* in 2015, it became clear that high configurational entropy can stabilize complex, otherwise immiscible oxide systems into single-phase structures at elevated temperatures [13]. The extreme compositional disorder intrinsic to HEOs disrupts conventional defect association trends, promoting new forms of electronic, structural, and ionic behavior [14]. This strategy has successfully been extended to lanthanide oxide systems, where the combination of cation diversity, aliovalent chemistry, and redox activity produce a rich landscape of metastable structures and defect chemistry with relevance for solid-state ionics, catalysis, and electronic devices [15–19].

Lanthanide high-entropy oxides (LN-HEOs) commonly adopt either disordered fluorite- or oxygen-deficient bixbyite-derived structures, with phase selection governed by average cation valence, coordination preferences, oxygen vacancy concentration, and synthesis history [15,16,18–21]. Fluorite structures, exemplified by ceria, $CeO_2$, favor tetravalent cations and exhibit high oxygen ionic conductivity due to their well-known ability to accommodate disordered oxygen vacancies. In contrast, bixbyite structures stabilize trivalent lanthanide cations in distorted sixfold coordination and feature ordered oxygen vacancy sublattices, which reduce ionic transport but lower the system enthalpy at high vacancy concentrations. The presence of multivalent cations such as Ce and Pr introduces further complexity by enabling mixed-valence behavior ($Ce^{4+}/Ce^{3+}$, $Pr^{4+}/Pr^{3+}$), defect-driven transport, and structural transitions under varying thermal or chemical environments.

A prototypical example of this behavior is the equimolar LN-HEO $(CeLaPrSmY)O_{2-\delta}$, hereafter referred to as 'flexite' for its variable fluorite-bixbyite character [18,19]. As illustrated in Figure 1, the ideal fluorite phase exhibits a highly symmetric cation and anion sublattice. In contrast, the bixbyite phase features a doubled unit cell with 25% ordered oxygen vacancies, lowering enthalpy at high vacancy concentrations while simultaneously limiting ionic transport. Flexite's ability to adopt either of these structures highlights the energetic proximity of multiple metastable states and underscores the importance of both configurational and electronic entropy in phase behavior. Prior experimental studies have demonstrated that Ce concentration plays a decisive role in determining phase stability in the $Ce_x(YLaPrSm)_{1-x}O_{2-\delta}$ system. Reducing Ce content often leads to bixbyite-dominated or multiphase materials [16,19,21]. These observations point to a delicate interplay between thermodynamic driving forces, redox chemistry, and kinetic trapping during synthesis, complicating efforts to rationally control phase selection.

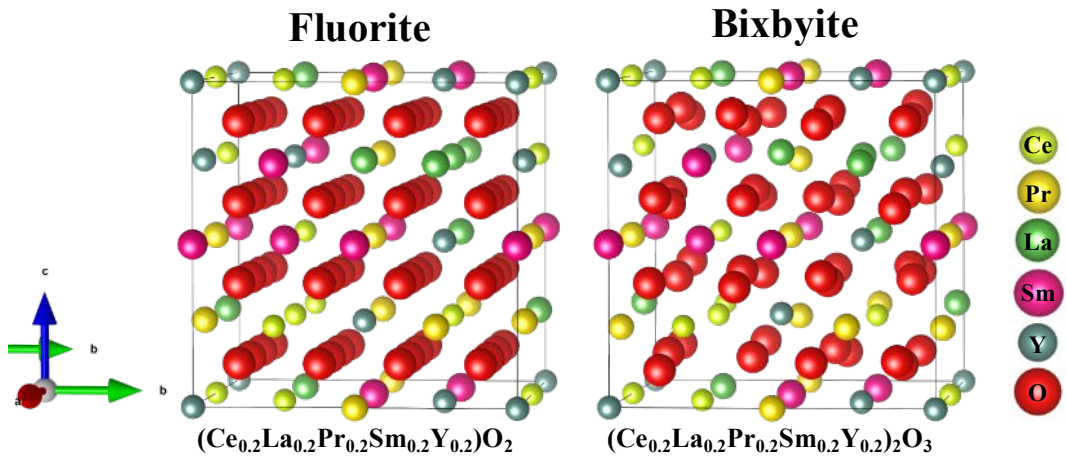

Figure 1. Idealized 2x2x2 fluorite supercell (left) and conventional bixbyite unit cell (right) for equimolar flexite in fluorite and bixbyite phases.

Despite growing experimental interest in LN-HEOs, several of their structure-property relationships remain unresolved. Prior studies have begun quantifying the roles of cation redox, oxygen vacancy formation, and configurational disorder in stabilizing fluorite-derived phases [16–19,21–26]; however, the relative importance of these contributions and their interplay across composition and vacancy concentration are not yet fully established. While oxygen vacancy concentration and ordering have been shown to influence phase stability and transport, there still lacks a systematic assessment of how vacancy concentration and spatial distribution affects lattice distortion, thermodynamic stability, and electronic structure across both fluorite and bixbyite polymorphs. The role of internal redox processes involving multivalent cations such as Ce and Pr further complicates this picture, as redox activity can couple to both local structure and electronic properties in nontrivial ways, affecting conductivity, charge compensation, and the evolution of the bandgap. At the same time, the highly disordered anion sublattice characteristic of LN-HEOs raises important questions regarding how structural complexity and configurational entropy may be deliberately leveraged to tune functional properties such as ionic conductivity or resistive switching. Addressing these coupled effects is therefore critical for developing predictive design principles for functional LN-HEOs.

In this work, we address these challenges through a density functional theory (DFT) exploration of the equimolar flexite and related $Ce_x(YLaPrSm)_{1-x}O_{2-\delta}$ compositions in both its fluorite and bixbyite structural forms. By explicitly varying oxygen vacancy concentrations from 0% to 25% and tuning Ce content, we examine how vacancy distribution, cation redox behavior, and configurational entropy influence lattice parameters, local structural distortions, electronic structure, and thermodynamic stability. Bader charge and electronic structure analyses are used to disentangle the roles of internal redox and vacancy accommodation, while free-energy considerations reveal the extent to which entropy stabilizes disordered fluorite phases relative to vacancy-ordered bixbyite. Together, these results provide a unified mechanistic framework for understanding phase stability and defect-driven functionality in LN-HEOs, offering guidance for the design of next-generation high-entropy oxide electrolytes and redox-active materials.

## 2. METHODS
### 2.1. Computational details
First-principles DFT calculations were executed using version 6.3.0 of the Vienna ab initio Simulation Package (VASP) [27,28] software, utilizing the projector augmented-wave (PAW) method [29]. The r$^2$SCAN [30] meta-GGA exchange-correlation functional was employed. The r$^2$SCAN functional has shown strong performance for describing highly correlated electron systems, such as lanthanide oxides, providing improved localization of $f$-electrons relative to standard GGAs [31]. The pseudopotential's atomic core electrons for Ce, La, Pr, Sm, Y, and O were treated with the Ce, La, Pr, Sm, Y_sv, and O PAW 64 datasets, where the $f$ electrons of Ce, Pr, and Sm are explicitly treated as valence electrons. The plane wave kinetic energy cutoff was set to 700 eV with a convergence threshold of $10^{-6}$ eV for each self-consistent electronic calculation, and the ionic relaxations iterated until the atomic forces were less than 0.02 eV/Å. Γ-centered k-point grids were generated automatically by using KSPACING values of either 0.4 $\text{Å}^{-1}$ for oxides or 0.15 $\text{Å}^{-1}$ for metals.

For the density of states (DOS) and band gap calculations presented in section 3.3.2, a r$^2$SCAN+$U$+SOC approach was utilized. An onsite Hubbard-type ($U$) parameter was applied to the *f*-states (*d*-states for La) through the simplified rotationally invariant framework developed by Dudarev *et. al*. [32]. Consistent with our prior validation of lanthanide oxides [31], the same Hubbard $U$ values were applied. Specifically, $U$ values of 0, 1, 4, 2.5, 7, and 0 were used for Y, La, Ce, Pr, Sm, and O, respectively. SOC calculations were initialized with the spin quantization axis along the z-direction (SAXIS = 0 0 1) and the moments were allowed to self-consistently solve. Atomic magnetic moments were initialized in a ferromagnetic configuration for both collinear and non-collinear calculations.

Atomic charges were quantified via Bader charge analysis. The Bader atomic volumes were partitioned using the program developed by Henkelman and co-workers using the total electron density [33–35]. The valence charge densities were integrated over these regions to obtain the Bader atomic charges. The analysis applied the approximate all-electron charge density derived from combining the AECCAR and CHGCAR files generated by VASP. The Bader net ionic charges are calculated by subtracting the Bader projected valence charges from the valence electron count in the PAW potentials.

To model the random oxides, seven 96-atom fluorite supercells ($Fm\overline{3}m$) were generated using size 2×2×2 conventional fluorite unit cells of $Ce_x(YLaPrSm)_{1-x}O_2$ with Ce concentrations x = 0.22, 0.31, 0.34, 0.38, 0.50, 0.81, and 0.88. The cation sublattice, consisting of 32 atoms, was decorated using the special quasi-random structure (SQS) algorithm [36] via the integrated cluster expansion toolkit (ICET) [37]. Oxygen vacancies were treated as explicitly vacated sites on the fluorite anion sublattice. For each Ce concentration, identical cation configurations were constructed to generate the corresponding bixbyite ($Ia\overline{3}$) supercells, which also contain 32 cations per 1×1×1 conventional unit cell. For each $Ce_x(YLaPrSm)_{1-x}O_{2-\delta}$ composition, several oxygen non-stoichiometries δ were considered. In fluorite structures, oxygen vacancies $V_O$ were introduced randomly, whereas in bixbyite structures the oxygen vacancies were restricted to the Wyckoff *8b* positions, in accordance with the crystallographic $Ia\bar{3}$ site symmetry. Except for 0% $V_O$, 25% $V_O$, and 30% $V_O$, each oxygen deficient composition has at least two distinct anion configurations to capture effects of local vacancy arrangement. At 25% $V_O$, the equimolar composition (flexite) had two fluorite anion configurations, while all other compositions used a single 25% $V_O$ configuration. For additional comparisons, calculations were performed for single-cation $CeO_{2-\delta}$ and $PrO_{2-\delta}$. All considered $Ce_x(YLaPrSm)_{1-x}O_{2-\delta}$ compositions are listed in Table S.1.

## 2.2. Thermodynamic equations

The well-known Gibb's free energy relation, $\Delta G = \Delta H - T\Delta S$, is the basis of thermodynamics for solid-state solutions. By calculating $\Delta G$ for the systems under consideration here, we extend the DFT calculation results to include temperature dependence and make thermodynamically relevant interpretations.

The formation enthalpy $\Delta H_f$ was calculated from the DFT total energies, and its equation is taken as the energy released when a LN-HEO compound is formed from the elemental constituents in their standard states:

$$\Delta H_f^{298K} \approx \Delta H_f^{0K} = E_{tot} - \sum_{x_i} x_i \mu_i \qquad (1)$$

where $\Delta H_f^{0K}$ is the formation energy calculated by DFT, $E_{tot}$ is the total energy of the LN-HEO containing $x_i$ atoms of element *i*, which has an elemental chemical potential of $\mu_i$ per atom. The chosen elemental reference states are listed in Table S.2.

In general, the total solid-state entropy $\Delta S_{tot}$ is a sum of configurational, vibrational, and electronic (i.e. magnetic, orbital, etc.) entropies [38,39]:

$$S_{tot} = S_{config} + S_{vib} + S_{el} + S_{mag} + S_{orb} + S_{charge} + \ldots \qquad (2)$$

Of these various entropic contributions in non-stoichiometric oxides, the largest is the configurational entropy associated with distributing oxygen vacancies $V_O$ across the anion sublattice, as well as the arrangement of charge-compensating cations, e.g., localized electrons, on the cation sublattice [40,41]. Here, we assume with ideal, uniform mixing that the largest difference in total entropy between the fluorite and bixbyite phases will be $\Delta S_{config}$, specifically from the anion sublattice $\Delta S_{config,anion}$, where $\Delta S_{config} = \Delta S_{config,cation} + \Delta S_{config,anion}$. Thus, we consider the estimated Gibbs free energy of formation at temperature $T$ to be

$$\Delta G_f = \Delta H_f - T\Delta S_{config}. \tag{3}$$

The configurational entropy per cation site, using Stirling's approximation, is

$$\Delta S_{config,cation} \approx -k_B \sum_i x_i \ln(x_i) \tag{4}$$

where $x_i$ is the relative fractions of each cation $i$. We define $\Delta S_{config,anion}$ as

$$\Delta S_{config,anion} = k_B \ln\left(\frac{N_O!}{n!(N_O - n)!}\right) \tag{5}$$

where $N_O$ is the number of oxygen sublattice sites, $n$ is the number of vacancies, and $k_B$ is the Boltzmann constant [40–42]. Using Stirling's approximation $\ln(a!) \approx a\ln(a) - a$, the entropy change associated with the formation of $n$ vacancies from Equation (5) becomes

$$\Delta S_{config,anion} = -N_O k_B \left(\frac{n}{N_O}\ln\left(\frac{n}{N_O}\right) + \frac{N_O - n}{N_O}\ln\left(\frac{N_O - n}{N_O}\right)\right). \tag{6}$$

Since bixbyite has ordered vacancies, we assume that it has no configurational entropy associated with distributing oxygen vacancies randomly across the anion sublattice, i.e. $\Delta S_{config,anion} = 0$.

A fully rigorous thermodynamic treatment of HEOs would require the inclusion of additional effects beyond those considered here, such as strain energy, oxygen chemical potentials, phase decomposition energetics, finite-temperature heat-capacity contributions, and more advanced treatments of configurational entropy [43–45]. While incorporating these effects could improve quantitative predictions of phase stability, developing such a comprehensive framework for chemically complex systems remains challenging. Within these limitations, our results show that treating cation and anion configurational entropy at the ideal-mixing level nonetheless provides a reasonable first-order description of the thermodynamic trends governing phase stability in this system.

## 3. RESULTS AND DISCUSSION

### 3.1. Structures as a function of $V_O$ and Ce content

#### *3.1.1 Lattice parameters*

In Ce-rich LN-HEOs under ambient $pO_2$, $Ce^{4+}$ is thermodynamically favored [19,21], requiring charge compensation through oxygen vacancy formation associated primarily with the trivalent lanthanide cations. In this limit, the oxygen vacancies occupy sites on the anion sublattice and are treated as vacated oxygen sites, which may be either randomly distributed (disordered defective fluorite) or spatially correlated (bixbyite-like ordering). Hereafter, references to oxygen vacancies correspond to explicitly vacated sites on the anion sublattice as described in Methods.

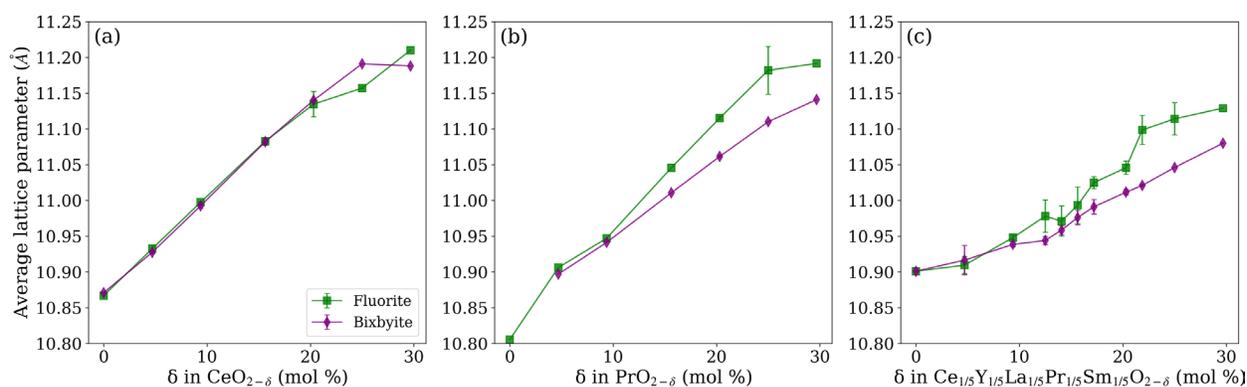

Figure 2. Computed supercell lattice constants of fluorite and bixbyite (a) $CeO_{2-\delta}$, (b) $PrO_{2-\delta}$, and (c) LN-HEO phases as a function of oxygen vacancy concentration. Error bars are the standard deviation between calculations.

Oxygen vacancy formation can lead to significant changes in lattice unit cell volumes due to a combination of electrostatic effects and cation redox responses. Figure 2 illustrates the evolution of lattice parameters for fluorite and bixbyite $CeO_{2-\delta}$, $PrO_{2-\delta}$, and equimolar flexite $(Y_{1/5}La_{1/5}Ce_{1/5}Pr_{1/5}Sm_{1/5})O_{2-\delta}$ structures as a function of oxygen vacancy concentration. It should be noted that the lattice parameter of fluorite is half that of bixbyite, but for comparison the lattice parameter of the supercell is plotted here. Overall, in Figure 2(c) the LN-HEO bixbyite phase exhibits systematically smaller lattice constants than the disordered fluorite phase at comparable vacancy concentrations with the same trend indicated in Figure S.1 for Ce concentrations up to 50% in $Ce_x(YLaPrSm)_{1-x}O_{2-\delta}$. This behavior is consistent with the expectation that random placement of unoccupied oxygen lattice sites in the fluorite structure induces greater local Coulombic repulsions, leading to larger lattice expansions compared to the more ordered vacancy arrangement in bixbyite. At low vacancy levels (≤ 10% Vo), the fluorite and bixbyite LN-HEO phases exhibit similar lattice constants, reflecting comparable structural responses to sparse vacancies.

When compared to $CeO_{2-\delta}$ [Figure 2(a)] and to compositions with Ce concentrations over ~81% in $Ce_x(YLaPrSm)_{1-x}O_{2-\delta}$ [Figure S.1], the fluorite and bixbyite lattice parameters of the supercells remain nearly identical across increasing oxygen vacancy concentrations, increasing at comparable rates with more pronounced divergence in the high-entropy compositions below 80% Ce. This suggests that additional factors other than Coulomb repulsion influence the relatively larger lattice expansion in the LN-HEO fluorite phases. In contrast, the trend in the average lattice parameter of $PrO_{2-\delta}$ [Figure 2(b)] parallels that of flexite [Figure 2(c)] and up to Ce 50% in $Ce_x(YLaPrSm)_{1-x}O_{2-\delta}$ [Figure S.1], where beyond approximately 10% $V_O$, the fluorite phase exhibits a more pronounced lattice expansion relative to bixbyite. At Ce concentrations larger than 50%, the percentage of Pr is less than 10% and is not expected to greatly contribute to the expanded lattice.

This behavior in fluorite structured $Ce_x(YLaPrSm)_{1-x}O_{2-\delta}$, likely reflects competing effects between vacancy-driven structural relaxation and changes in cation oxidation states, where the increasing presence of oxygen vacancies promotes the reduction of $Pr^{4+}$ to $Pr^{3+}$. The larger ionic radius of $Pr^{3+}$ contributes to lattice expansion, which becomes particularly pronounced between 14% and 20% Vo (discussed in Section 3.3.1). At the same time, random vacancy placement leads to local environments that give greater variability in between regions that are contracted or expanded. Conversely, the bixbyite phase shows a smoother, nearly linear increase in lattice expansion with vacancy concentration. The ordered vacancy arrangement in bixbyite minimizes local strain accumulation, allowing a more homogeneous structural response to increasing oxygen deficiency. Once the cation sublattice becomes more homogenous at Ce content greater than 50%, fluorite also experiences more uniformed structural response regardless of random placement of Vo.

### 3.1.2. Local structural distortion

To understand how oxygen vacancies and internal redox drive structural heterogeneity in LN-HEOs, we analyzed deviations in local bonding environments, specifically LN–O bond lengths, coordination numbers, and polyhedral distortions, as a function of oxygen vacancy concentration δ for both fluorite and bixbyite phases. Figure 3 displays the average LN–O bond lengths for each LN element as a function of δ in equimolar fluorite and bixbyite LN-HEO structures, with the distribution of those bond lengths shown in Figure S.2 and Figure S.3, respectively. Each point in Figure 3 represents the mean bond length across multiple supercells.

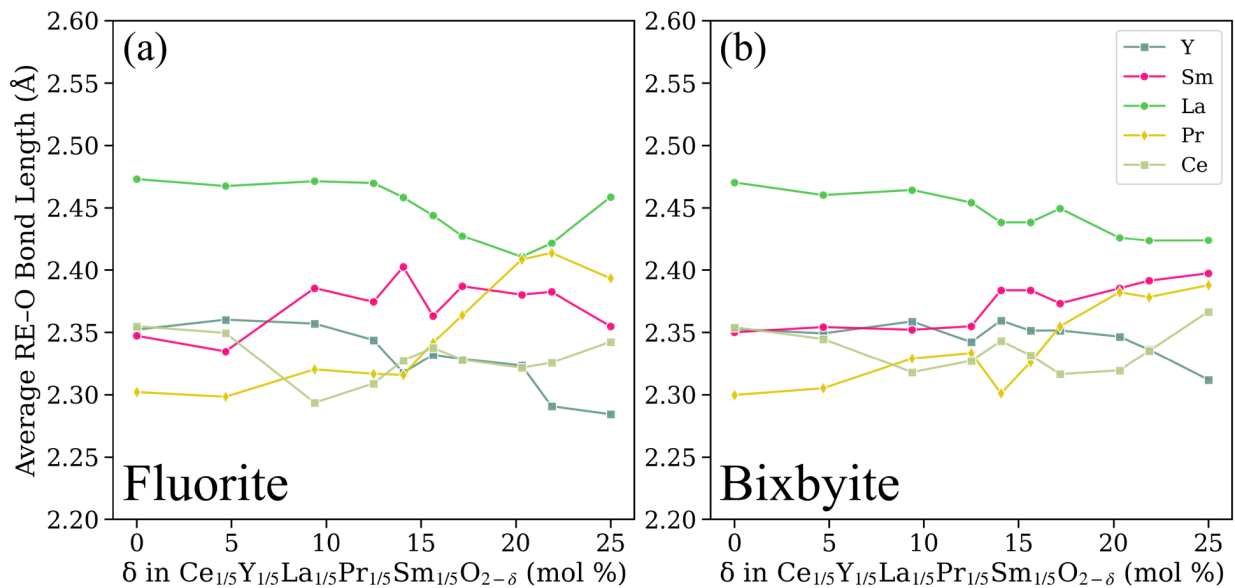

Figure 3. Average LN–O bond lengths (in Å) for flexite in fluorite (left) and bixbyite (right) $(Ce_{1/5}Y_{1/5}La_{1/5}Pr_{1/5}Sm_{1/5})O_{2-\delta}$ structures as a function of oxygen non-stoichiometry (δ). Each point represents the mean bond length across multiple supercells, capturing the effect of vacancy-driven local structural distortions and cation-specific redox behavior on the average lanthanide coordination environment. Lines connecting points are to guide the eye.

At low vacancy concentrations (δ ≤ 5%), the LN–O bond lengths of Ce, Sm, and Y are relatively uniform, clustering around 2.35 Å, reflecting the nearly symmetric coordination environments [Figure S.4, Figure S.5]. In this low vacancy region, the fluorite and bixbyite phases exhibit similar average bond lengths, with minor variations due to ionic size differences (e.g., $La^{3+}$ > $Pr^{3+}$ > $Sm^{3+}$ > $Y^{3+}$ > $Pr^{4+}$ > $Ce^{4+}$). However, as vacancies are introduced, bond lengths begin to diverge, especially in the disordered fluorite phase. The LN–O bond length distributions [Figure S.2, Figure S.3] and lanthanide coordination number distributions [Figure S.4, Figure S.5] confirm that both phases exhibit broadening and asymmetry in LN–O environments as δ increases, though the effect is more pronounced in fluorite. Unlike fluorite, the bixbyite LN–O bond length distributions begin to narrow once the oxygen vacancy concentration surpasses ~20% $V_O$. This highlights the increasing effectiveness of vacancy ordering at high oxygen non-stoichiometries, where $V_O$ occupy crystallographically defined sites in the bixbyite structure. At lower vacancy concentrations, bixbyite contains relatively few vacancies, and their local structural influence is limited, resulting in bond-length distributions comparable to those of disordered fluorite. As the vacancy concentration increases toward the stoichiometric 25% $V_O$ limit of bixbyite, cooperative vacancy ordering reduces local strain and Coulombic repulsion, leading to a more uniform coordination environment and narrower LN–O bond-length distributions.

In the ordered bixbyite phase, LN–O bond lengths show only slightly less variation than fluorite across most δ, with more significant narrowing of the distributed LN–O bonds observed at the highest δ values [Figure S.3]. This suggests that while the ordered vacancy sublattice in bixbyite does impose some structural regularity, it is not sufficient to fully suppress local distortions at intermediate vacancy levels. These results support the idea that vacancy disorder amplifies local strain and LN–O variability, while vacancy ordering in bixbyite partially suppresses such distortions at high defect concentrations. The modest difference between fluorite and bixbyite highlights how configurational disorder and internal redox in HEOs can introduce significant local strain, even in nominally ordered structures. This also underscores the limits of vacancy ordering in fully stabilizing the local structure at moderate defect concentrations.

Figure 3 captures the effect of vacancy-driven local structural distortions and cation-specific redox behavior on the average lanthanide coordination environment. A pronounced elongation of Pr–O bond lengths emerges near 15% Vo in both phases, coinciding with the onset of $Pr^{3+}/Pr^{4+}$ mixed valency [Figure 10]. This same trend is found across all Ce content [Figure S.6, Figure S.7] with the pronounced elongation occurring closer to 10% $V_O$ at higher Ce content. This is consistent with the larger ionic radius of $Pr^{3+}$ (ionic radius ~1.13 Å) compared to $Pr^{4+}$ (~0.96 Å), and highlights Pr's unique redox flexibility among the LN elements. Simultaneously, La–O bond lengths in Figure 3(a) exhibit a decrease at higher oxygen vacancies. This behavior is attributed to a redistribution of local strain, rather than redox behavior since La valency does not appreciably change with increasing $V_O$ [Figure 9, Figure S.13, Figure S.14]. Instead, the La–O dip in Figure 3(a) is a consequence of the sudden decrease in coordination number from the random placement of the $V_O$ [Figure S.4(d)] in the SQS. Across all $Ce_x$ concentrations, La has a high correlation between the amount of near-neighbor $V_O$ and the La–O bond length. To a lesser extent, the other mono-valent LN elements, such as Y, also exhibit correlation between coordination number and LN–O bond length. Yet, La is less resilient to the compression exerted by local strain, leading to La–O bonds contracting more strongly, likely due to La's higher compressibility [46].

To further quantify geometric asymmetry, we computed the polyhedral distortion index ($\Delta d$) for each LN-centered coordination polyhedron. An illustrative example of a 6 coordinated LN-centered polyhedron is shown in Figure 4(a). The polyhedral distortion index $\Delta d$ was calculated using the equation [6,47–49]:

$$\Delta d = \frac{1}{N_{CN}} \sum_{i=1}^{N_{CN}} \left( \frac{d_i - d_{av}}{d_{av}} \right)^2 \quad (7)$$

where $d_i$ is the individual LN–O bond length and $d_{av}$ is the average bond length of the polyhedron surrounding a LN with $N_{CN}$ oxygen coordinated number of bonds. This metric captures the deviation of individual LN–O bond lengths from the average, serving as a measure of local geometric distortions. As shown in Figure 4(a), fluorite exhibits a monotonic increase in $\Delta d$ with δ from the growing structural disorder driven by randomly distributed oxygen vacancies and mixed valence LN species. In contrast, $\Delta d$ in bixbyite [Figure 4(b)] increases at intermediate vacancy levels but then begins to decrease at approximately δ > 15%, where vacancy ordering imposes structural regularity. This same trend is found across Ce content [Figure S.8]. In Figure 4(c) and Figure 4(d), the largest $\Delta d$ values tend to be Ce and Pr in both fluorite and bixbyite. Trivalent-only LN cations (e.g., $La^{3+}$, $Sm^{3+}$, $Y^{3+}$) show lower $\Delta d$ values, lacking the redox flexibility to accommodate local changes in coordination or charge.

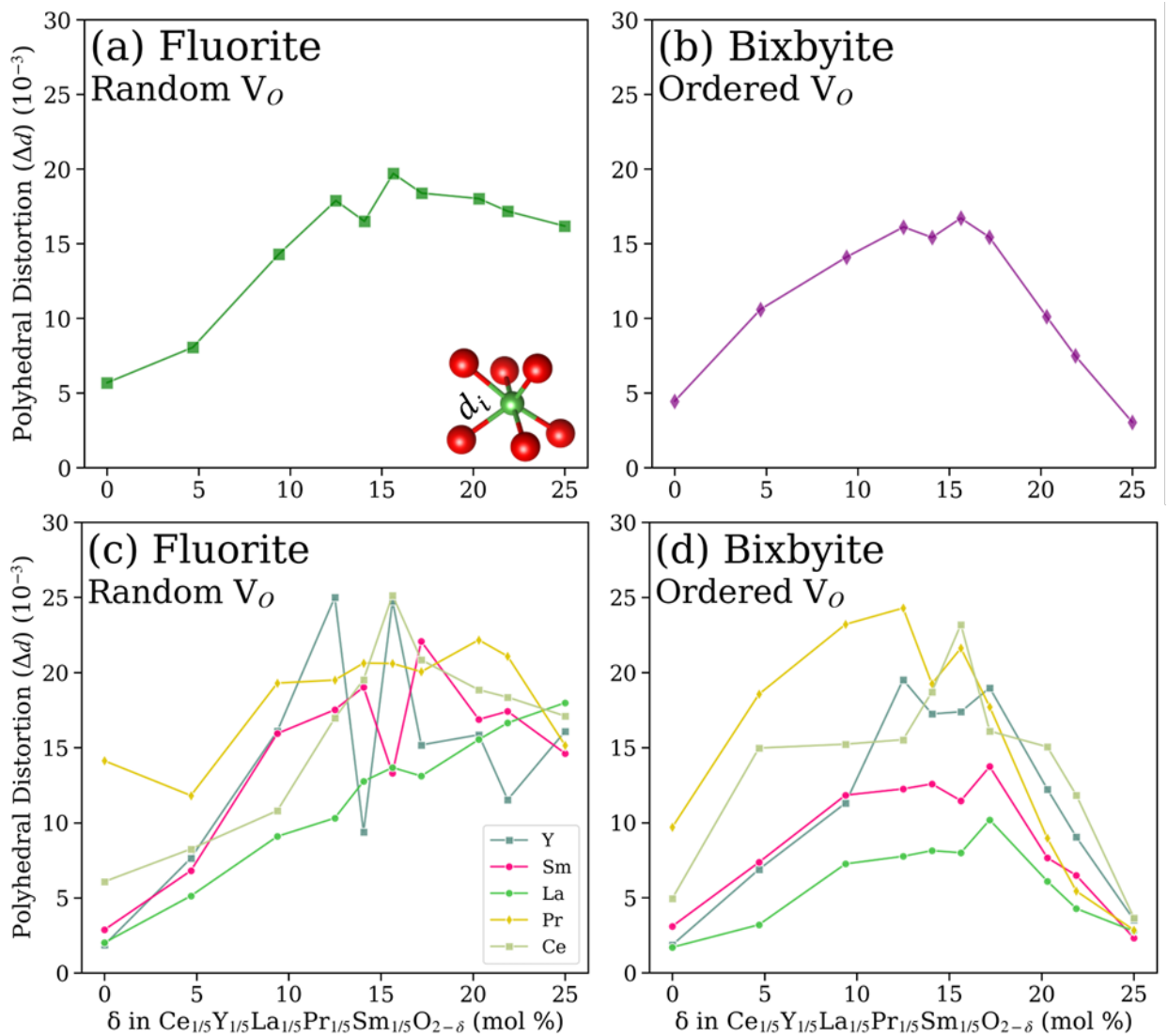

Figure 4. Average polyhedral distortion index ($\Delta d$) for LN-centered coordination environments in (a, c) fluorite and (b, d) bixbyite structures as a function of oxygen vacancy concentration. Panels (a) and (b) show $\Delta d$ values averaged over all LN–O polyhedral, while (c) and (d) present the $\Delta d$ contributions from induvial LN cations. The distortion index quantifies deviations from ideal polyhedral geometry, with an example LN centered coordination polyhedron illustrated in (a).

These results demonstrate that local structural distortions in LN-HEOs are strongly modulated by both oxygen vacancy concentration, oxygen vacancy ordering, and cation chemistry. The disordered fluorite structure shows greater heterogeneity in LN–O bonding and polyhedral distortion, especially in the presence of redox-active cations like Pr and Ce. These types of distortions influence macroscopic properties such as thermal conductivity and ionic transport, emphasizing the importance of understanding local bonding environments in HEOs.

## 3.2. Thermodynamic stability and formation energy

The formation enthalpy, $\Delta H_f$, is a fundamental thermodynamic descriptor that reflects the energetic favorability of compound formation from its constituent elements. Using $\Delta H_f$ within a convex hull framework allows assessment of the relative stability of competing solid-state compositions [50]. Previous work established the r$^2$SCAN exchange-correlation functional provides reliable $\Delta H_f$ predictions for lanthanide binary oxides, with a mean relative error of less than 3% when compared to room-temperature experimental values [31]. Leveraging this accuracy, we now use $\Delta H_f$ via the convex hull formalism to assess the relative stability of our equimolar LN-HEO system with varying concentrations and configurations of oxygen vacancies, $V_O$. Figure 5 displays the minimum computed formation energies per atom $\Delta H_f$ for $CeO_{2-\delta}$, $PrO_{2-\delta}$, and flexite $(Ce_{1/5}Y_{1/5}La_{1/5}Pr_{1/5}Sm_{1/5})O_{2-\delta}$ as a function of $\delta$ concentration for both fluorite (random $V_O$ anion sublattice) and bixbyite (ordered $V_O$ anion sublattice), enabling comparison of vacancy-driven phase stability between the single-cation component and multicomponent limits.

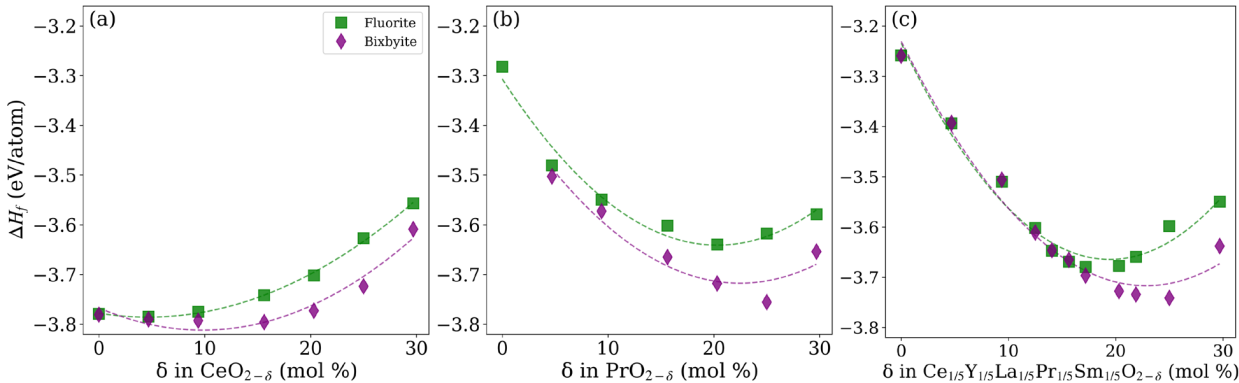

Figure 5. Predicted minimum formation enthalpy $\Delta H_f$ (in eV/atom) of (a) $CeO_{2-\delta}$, (b) $PrO_{2-\delta}$, and (c) $Ce_{1/5}Y_{1/5}La_{1/5}Pr_{1/5}Sm_{1/5}O_{2-\delta}$ as a function of oxygen vacancy concentration. Green squares represent fluorite structure and purple diamonds represent bixbyite structures. Each point corresponds to the lowest-energy configuration at a given $\delta$, extracted from a set of DFT calculations across distinct anion arrangements (see Figure S.9). The dashed lines are the second-order least-squares polynomial fit.

Across the vacancy concentration range in the LN-HEO shown in Figure 5(c), we predict a second-order nonlinear dependence of $\Delta H_f$, underscoring the complex interplay between vacancy content and structural ordering. At vacancy concentrations below approximately 15%, fluorite and bixbyite phases exhibit nearly degenerate formation enthalpies, suggesting comparable thermodynamic stability irrespective of oxygen sublattice arrangement. However, distinct differences emerge at higher vacancy levels. The fluorite structure reaches a minimum $\Delta H_f$ between 17% and 20% Vo before becoming increasingly energetically unfavorable at higher δ. In contrast, the bixbyite structure maintains decreasing formation energies with increasing vacancy concentration up to its energetic minimum at 25% Vo. This reflects the growing enthalpic advantage of vacancy ordering in the bixbyite phase at high δ from the reduced Coulombic repulsion and local strain minimization via cooperative oxygen vacancy ordering, which aligns with the average polyhedral distortion presented in Section 3.1.2.

In comparing to the equimolar LN-HEO, the single-cation $CeO_{2-\delta}$ and $PrO_{2-\delta}$ systems show qualitatively similar second-order $\Delta H_f$ curves. Nonetheless, the fluorite-to-bixbyite crossover occurs at either significantly lower $V_o$ concentrations, around 5% $V_O$ for $CeO_{2-\delta}$, or is entirely absent in $PrO_{2-\delta}$. As the concentration of Ce increases in $Ce_x(YLaPrSm)_{1-x}O_{2-\delta}$, the $\Delta H_f$ progressively decreases in low vacancy region (below 10% $V_O$). Concurrently, at concentrations greater than 50% Ce, fluorite-to-bixbyite crossover presents below 10% $V_O$. These deviations from equimolar flexite indicate the unique stabilization effect of configurational entropy in LN-HEOs, which allows the disordered fluorite phase to remain thermodynamically competitive at higher vacancy concentrations.

To further evaluate the role of cerium in stabilizing fluorite or bixbyite phases, Figure 6 shows the computed convex hulls of minimum free energies $\Delta G_f$ per atom for $Ce_x(YLaPrSm)_{1-x}O_{2-\delta}$ compositions in both fluorite (random $V_o$) and bixbyite (ordered $V_o$) structures at various temperatures ($T$ = 0K to 1750K). For each Ce concentration, only the lowest-energy $\Delta G_f$ among the considered $V_O$ concentrations [Figure S.10] is plotted; these minima form the solid green (fluorite) and solid purple (bixbyite) convex-hull lines. At 0K, $\Delta H_f$ for fluorite decreases monotonically with increasing Ce content, reflecting the stabilizing effect of $Ce^{4+}$ within the fluorite environment. Whereas bixbyite remains enthalpically preferred over fluorite across all Ce concentrations. This contrasts with experiment, where fluorite is observed to emerge at Ce contents at as low as ~35% [19]. So, enthalpy alone does not explain the experimentally observed fluorite stability at lower Ce contents.

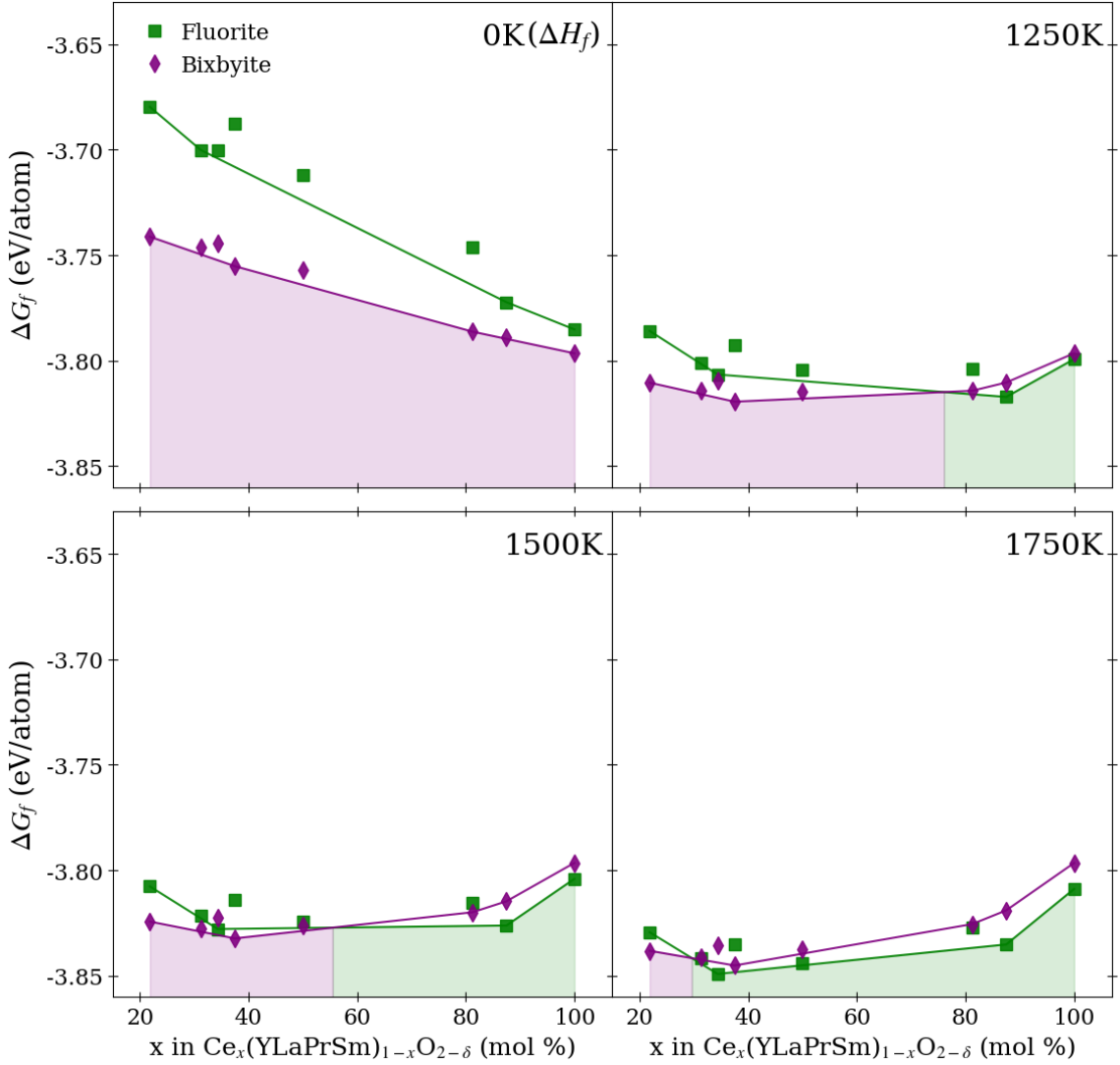

Figure 6. The Gibbs formation energy $\Delta G_f$ in eV/atom for fluorite (green squares) and bixbyite (purple diamonds) $Ce_x(YLaPrSm)_{1-x}O_{2-\delta}$ structures as a function of considered Ce content at T = 0K, 1250K, 1500K, and 1750K. Solid lines connect the lowest-energy points forming the convex hull for each phase with the shaded region indicating most favored phase.

When an estimate of finite-temperature effects is included through the ideal configurational entropy contributions $\Delta S_{config}$, a substantial stabilization of the fluorite phase is observed from the additional $\Delta S_{config,anion}$ contribution that the bixbyite phase lacks; noting that both fluorite and bixbyite structures have the same cation configurational entropy. We see that as we increase temperature in Figure 6, the Ce region in the fluorite stability window increases. The resulting free energies of formation $\Delta G_f$ at $T_{1750K}$, show fluorite becoming favored at ~30% Ce content. This predicted crossover in phase stability brings the theoretical transition much closer to the experimentally observed boundary, suggesting that entropy, particularly from disordered oxygen sublattices, plays a decisive role in stabilizing the high-temperature fluorite structure. Importantly, the overall $\Delta H_f$ and $\Delta G_f$ differences between the two phases across all Ce compositions remain modest (< 0.1 eV/atom), implying a thermodynamic near degeneracy that may allow kinetic and experimental processing factors to influence the observed phase.

The convex hull analysis in Figure 6 reveals that fluorite is not enthalpically preferred presumably because, although Ce improves fluorite stability, the enthalpic benefit of $Ce^{4+}$ cannot overcome bixbyite's ordered structural advantage in accommodating smaller trivalent cations with less local structural distortion and Coulombic repulsion. While the enthalpic stability $\Delta H_f$ favors the bixbyite phase, the inclusion of $\Delta S_{config,anion}$ alters the stability landscape, penalizing bixbyite's total $\Delta G_f$ from the lower entropic contribution of the ordered arrangement of oxygen vacancies. When the configurational entropic contribution is included in $\Delta G_f$, the fluorite structure becomes thermodynamically favored at increasingly lower amounts of Ce content. Since Ce remains predominantly in the 4+ oxidation state and Pr exhibits mixed-valent behavior regardless of fluorite or bixbyite phases [Figure 9, Figure 10], this shift is unlikely to be driven by redox behavior between phases, but by the increased configurational freedom in the disordered fluorite structure [21].

In order to enable an estimate of the thermodynamically preferred $V_O$ concentration at 0K [Figure 7(a)] and 1750K [Figure 7(b)], a second-order least-squares polynomial was fit to the lowest-energy configurations across varying $V_O$ levels for each Ce concentration [Figure S.11, Figure S.12]. The fits yield high coefficients of determination ($R^2 \approx 0.90$–$0.99$), validating the smooth and systematic variation of stability with oxygen vacancy content. We see that enthalpically [Figure 7(a)], bixbyite maintains higher oxygen vacancy content in comparison to fluorite, likely due to the reduction in Coulombic repulsion provided by the ordered vacancies. The fitted energetic minima for equimolar flexite occurs near 20% Vo for fluorite and 23% Vo for bixbyite. The enthalpic stabilization of vacancy rich concentrations near 20–25% may be further attributed, in part, to the internal redox behavior of multivalent cations, particularly Pr. As shown in Section 3.3.1, Pr readily adopts mixed-valent (3+/4+) states that seemingly become increasingly favorable as the system incorporates more oxygen vacancies. This redox adaptability provides an additional enthalpic driving force for both fluorite stabilization at intermediate vacancy levels and bixbyite ordering at high δ. When configurational entropy is included [Figure 7(b)], the lowest energy $V_O$ is higher, with fluorite getting closer to bixbyite levels of Vo %.

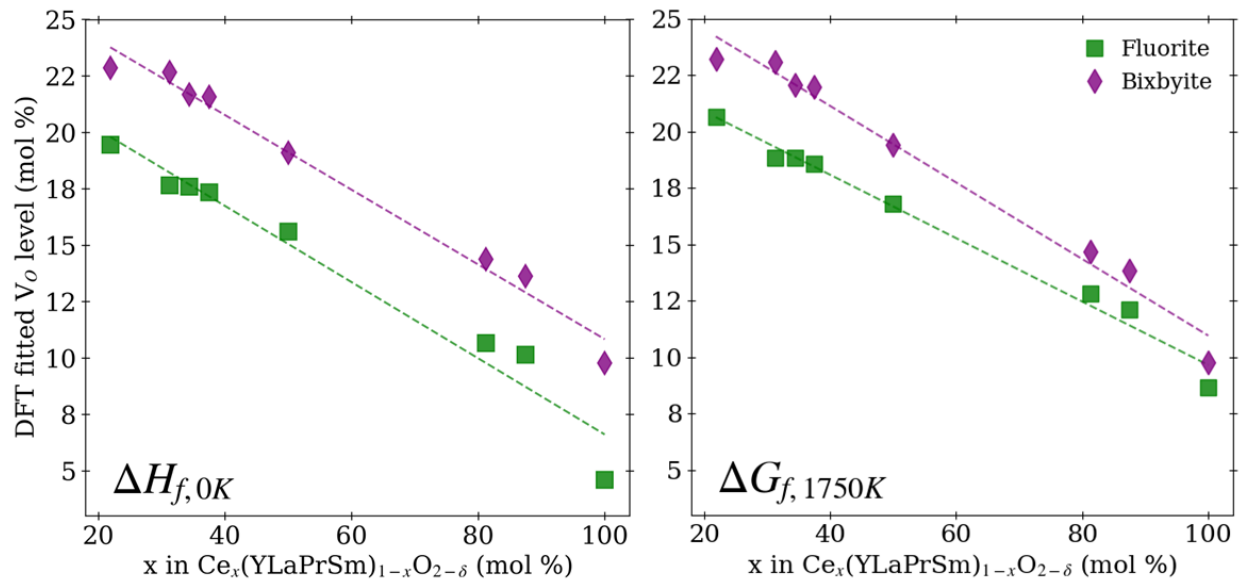

Figure 7. The predicted most favorable $V_O$ at each considered Ce concentration based on the energetic minimum from the second-order least squares polynomial fitting to DFT calculations of $\Delta G_f$ at (a) $T_{0K}$ and (b) $T_{1750K}$ for fluorite (green squares) and bixbyite (purple diamonds). Dashed lines are the least square linear fit.

While including vibrational entropy ($\Delta S_{vib}$) and considering short-range cation ordering may yield improved thermodynamic predictions, our results capture the primary thermodynamic drivers governing phase stabilization in LN-HEOs. This is justified given that the anion sublattice configurational entropy often dominates the entropy landscape in defective oxides, as discussed in Methods. Figure 8 presents the computed $\Delta H_f$ and $\Delta G_f$ of pristine fluorite (δ = 0, $Fm\bar{3}m$ symmetry), pristine bixbyite (δ = 0.5, $Ia\bar{3}$ symmetry), and highly oxygen deficient fluorite (δ = 0.5, $Fm\bar{3}m$ symmetry) compositions as a function of cerium concentration. In comparing pristine fluorite and pristine bixbyite, both structures lack anion disorder (i.e., $\Delta S_{config,anion} = 0$), so differences in $\Delta G_f$ arise primarily from cation configurational entropy $\Delta S_{config,cation}$ and enthalpic influences. Pristine fluorite (δ = 0.0) shows a linear decrease in both $\Delta H_f$ and $\Delta G_f$ with increasing Ce, highlighting its energetic preference for Ce-rich compositions. For bixbyite with 25% $V_O$, $\Delta H_f$ remains nearly constant across Ce concentrations, indicating that its enthalpic stability is largely insensitive to Ce content. The fluorite structure with 25% $V_O$ exhibits a similar behavior roughly parallel to the bixbyite curve. However, the fluorite line sits higher in formation energy. Since the major structural difference between these two compositions is the random placement of $V_O$ in fluorite versus the ordered $V_O$ arrangement in bixbyite, this offset reflects a clear enthalpic preference for vacancy ordering at 25% $V_O$. Reinforcing that at increasingly high Vo levels the entropic advantage of randomly distributed Vo in fluorite does not outweigh the enthalpic advantage of the vacancy-ordered bixbyite structure. Whereas their $\Delta G_f$ increases slightly with Ce, likely reflecting a reduced cation configurational entropy as the composition becomes less diverse. This underscores the critical role of Ce in stabilizing the disordered fluorite structure, in low oxygen vacancy concentrations. The contrasting behavior between the two phases highlights how both enthalpic and entropic factors jointly govern phase stability in this HEO system.

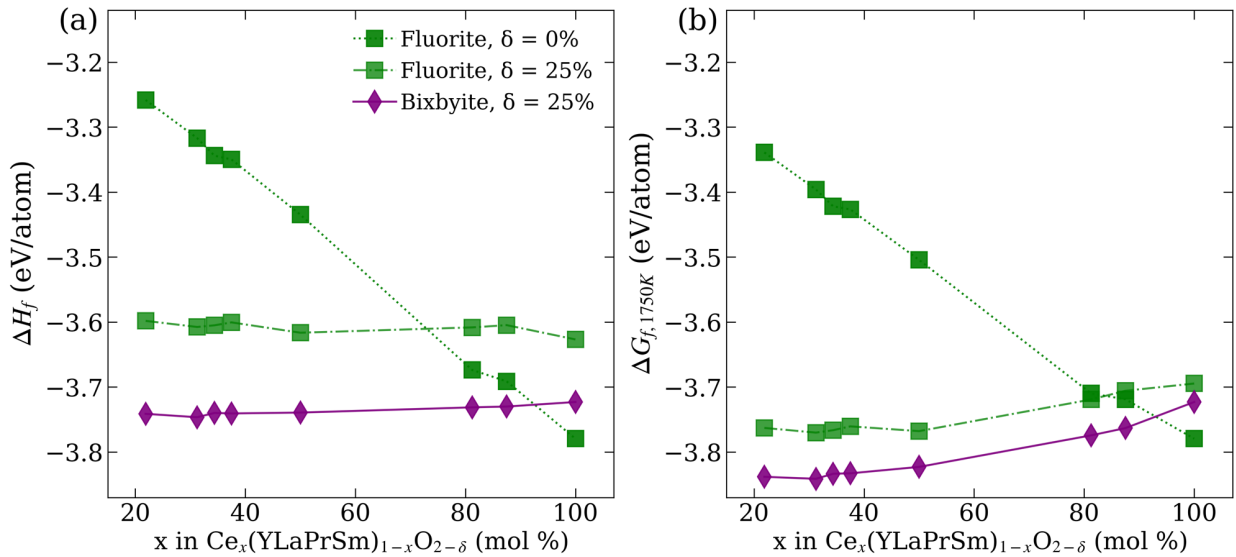

Figure 8. The computed (a) formation enthalpy $\Delta H_f$ and (b) Gibbs formation energy $\Delta G_f$ in eV/atom for fluorite with $Ce_x(YLaPrSm)_{1-x}O_2$ (dotted line), fluorite with $Ce_x(YLaPrSm)_{1-x}O_{1.5}$ (dot-dashed line), and bixbyite with $(Ce_x(YLaPrSm)_{1-x})_2O_3$ (solid line) compositions as a function of considered Ce content. Lines are to guide the eye.

Overall, these findings in Section 3.2 underscore the multifaceted nature of thermodynamic stability in Ce based LN-HEOs, where structural, configurational, and redox degrees of freedom are all interlinked. The convergence of these effects suggests that oxygen vacancy concentration not only dictates phase stability, but also modulates the electronic structure, highlighting the need for a coupled treatment of thermodynamics and electronic degrees of freedom, which we explore further in Section 3.3. Further, Ce's presence and the associated configuration entropy, particularly from anion disorder, are key to meta-stabilizing the fluorite phase. The inclusion of $\Delta S_{config,anion}$ shifts the enthalpically unfavored fluorite phase boundary to be energetically preferred over bixbyite at Ce content as low as 30% (based on $\Delta G_f$), in excellent agreement with experimental observations. This highlights the necessity for accounting for temperature-dependent entropy contributions when predicting phase stability in HEOs. The convex hull analysis indicates that although Ce incorporation reduces the energy penalty for the fluorite phase, this benefit is insufficient to overcome the inherent enthalpic advantage of bixbyite, whose ordered vacancy sublattice accommodates the smaller trivalent cations with reduced local distortion and lower Coulombic repulsion.

## 3.3. Electronic structure

### 3.3.1. Bader charges and valency

The number of *f*-electrons and the evolution of LN oxidation states play a critical role in interpreting the transport and electronic properties of oxide solid solutions [51]. In LN-based oxides, internal redox, namely, shifts in cation valency, can alter electron shell occupancy, affecting charge compensation, band alignment, and defect chemistry. This sensitivity to valence variation underscores the importance of resolving the oxidation behavior of each species across different structural and defect configurations. Previous experimental studies, determined that Ce predominantly exists in the 4+ state, Pr exhibits a mixed 4+/3+ valency, and the remaining cations (La, Sm, Y) consistently remain 3+ [15,16,23,25]. They further proposed that the observed bandgap narrowing is closely linked to the multivalent nature of Pr, reinforcing the importance of internal redox in tuning the electronic structure of LN-HEOs.

To investigate valency behavior across different oxygen vacancy concentrations, we evaluated Bader charges for each LN cation. These charges, computed without enforcing any formal oxidation constraints, reflect the self-consistent electronic redistribution of the system in response to varying δ. Atomic Bader charges were averaged for each element within the supercell and benchmarked against reference values derived from the corresponding binary oxides to infer oxidation trends. Figure 9 presents the average cation Bader charges for flexite across increasing δ values for both fluorite (right) and bixbyite (left) phases. The overall trends are similar across both structures, suggesting that the precise arrangement of oxygen vacancies exerts only a modest effect on average cation valency. Yet, more nuanced insight emerges when analyzing the distribution of individual cation charges. The individual Bader charge distributions for fluorite and bixbyite, presented in Figure S.13 and Figure S.14, respectively, show that the fluorite phase exhibits significantly larger charge variability at high δ, particularly for Ce and Pr, compared to bixbyite. This greater spread arises from the disordered nature of the fluorite structure, where local environments vary more substantially. La, Sm, and Y remain close to their expected 3+ states, with Sm exhibiting a slight reduction at higher δ, though remaining far from a 2+ state.

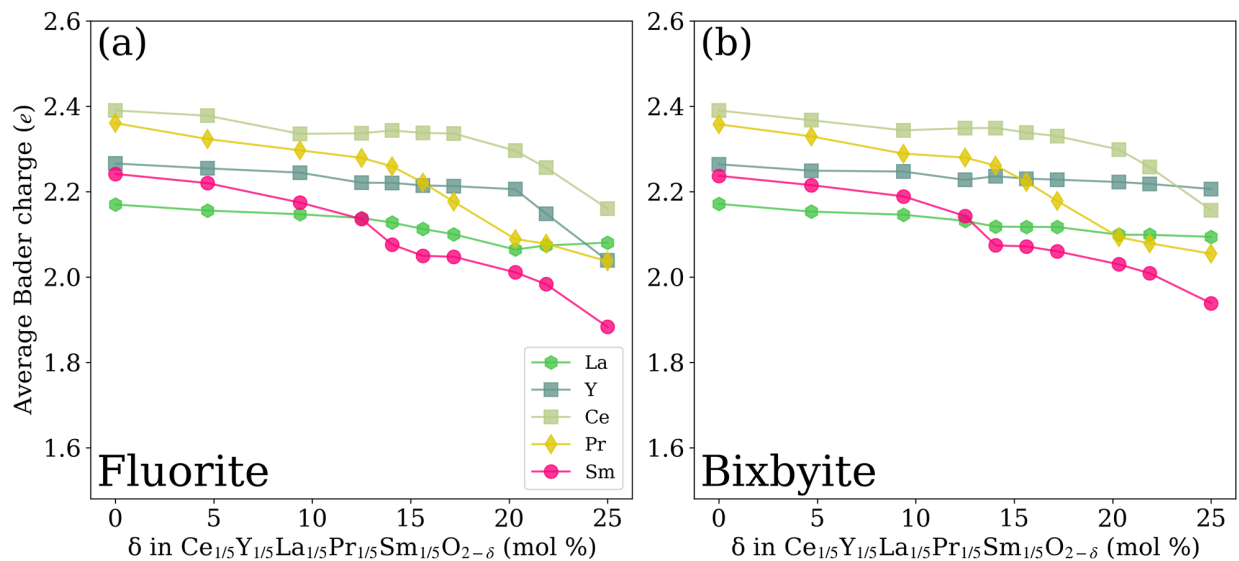

Figure 9. Average Bader charges of LN cations in $Ce_{1/5}Y_{1/5}La_{1/5}Pr_{1/5}Sm_{1/5}O_{2-\delta}$ as a function of oxygen vacancy concentration for fluorite (right) and bixbyite (left) structures. Each data point represents the mean Bader charge for a given element averaged across all supercell configurations. Lines are included to guide the eye.

Looking at Figure 9(a), the predicted drop in Y charge in the fluorite structure can be attributed to the random placement of vacancies. This anomaly is explained by examining the local atomic environments of the oxygen vacancies in the different SQS. For instance, the two 25% $V_O$ SQS configurations, one had 14 of its 16 oxygen vacancies with at least one yttrium in its 1NN local environment, while the other SQS had only 7 Y-associated vacancies. Consequently, the greater number of Y–$V_O$ associations in the first SQS led to a notable reduction in the average Y Bader charge. This reiterates how structure-property analysis can clarify outliers resulting from random vacancy generation within SQS models.

Figure 10 focuses on the distributions of Ce and Pr Bader charges as a function of δ within the LN-HEO. At low vacancy concentrations, both cations are predominantly 4+, but as δ increases, site specific reductions occur, broadening the charge distribution. Notably, at ~25% $V_O$, bixbyite exhibits a sharp collapse toward fully reduced 3+ states for both Ce and Pr, while fluorite retains a significant spread in Bader charges, indicating locally heterogenous reduction. This supports a vacancy-driven internal redox mechanism in both phases, with greater variability in disordered fluorite. Pr is the most redox-flexible of the LN cations, transitioning from mostly 4+ at low δ to a strongly mixed 3+/4+ state near δ ≈ 15%-20%, and to more 3+ at δ > 22-25% [Figure 10(b)]. In contrast, Ce reduction is markedly delayed in the high-entropy oxide relative to the single-cation system: whereas Ce in $CeO_{2-\delta}$ begins to deviate from the 4+ state at ~10% $V_O$ [Figure S.15], Ce in the LN-HEO remains largely tetravalent until substantially higher vacancy concentrations. This delayed reduction highlights the stabilizing effect of the high-entropy cation environment on $Ce^{4+}$.

These observations align with Section 3.2, where the enthalpic minima for both fluorite and bixbyite structures occur near δ ≈ 20%-25%, suggesting that redox transitions are intimately tied to thermodynamic stabilization. These results are consistent with prior experimental findings [15,16,21,23,25], confirming that Ce remains mostly 4+, Pr exhibits mixed valency, and La, Sm, and Y, are nominally trivalent across a wide range of oxygen vacancy concentrations.

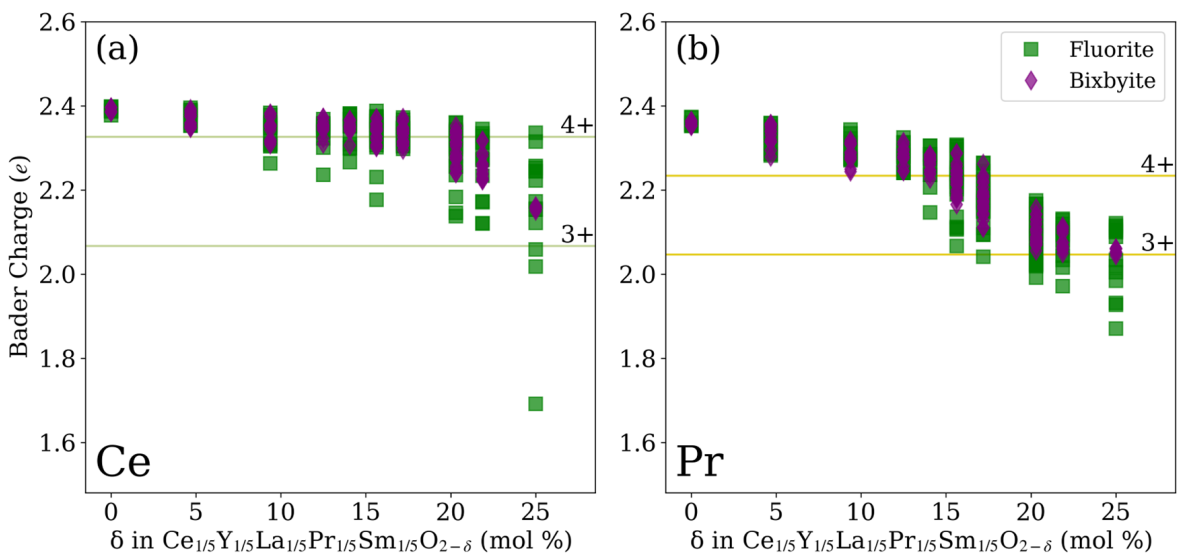

Figure 10. Distribution of individual LN-HEO Bader charges for (a) Ce and (b) Pr across increasing oxygen vacancies for fluorite (green squares) and bixbyite (purple diamonds) structures. Horizontal solid lines mark reference Bader charges from the corresponding binary oxides.

### 3.3.2. Density of states and band gaps

The electronic structure of LN-HEOs is highly sensitive to both oxygen stoichiometry and internal redox processes, particularly those involving *f*-electron redistribution in multivalent cations like

Ce and Pr. DFT calculations provide insight into how these redox transitions modulate the band gap and density of states (DOS) across varying $V_O$ concentrations. Figure 11 shows the average DFT-predicted minimum electronic band gaps for fluorite and bixbyite phases of $Ce_{1/5}Y_{1/5}La_{1/5}Pr_{1/5}Sm_{1/5}O_{2-\delta}$ as a function of δ (see also Figure S.16). Band gap values span from approximately 0.2eV to 1.6eV depending on the vacancy concentration, with bixbyite exhibiting slightly higher gaps than fluorite between 5%-14% $V_O$ and again beyond 20% $V_O$. A notable feature in both structures is a sharp increase in band gap between 14% (~0.6 eV) and 15% (~1.2 eV) $V_O$, followed by a drop between 15% and 20% $V_O$ (~0.6 eV). This nonmonotonic behavior reflects a transition in the electronic structure that coincides with the onset of $Pr^{3+}$ formation.

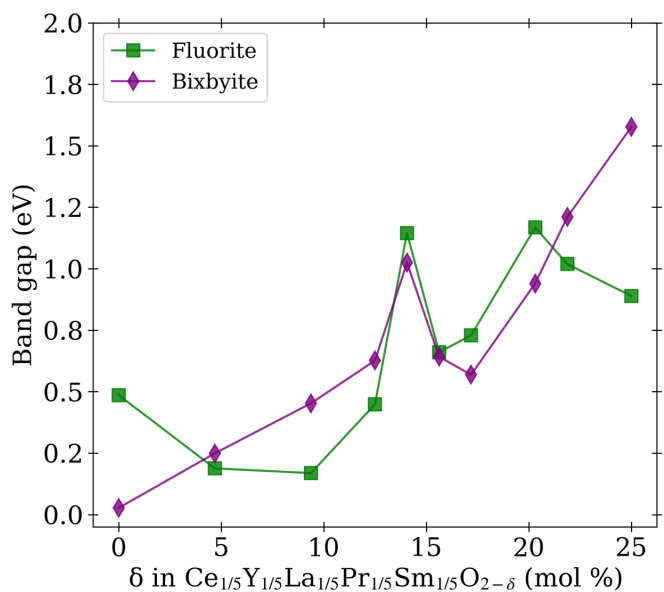

Figure 11. DFT predicted band gap (in eV) of $Ce_{1/5}Y_{1/5}La_{1/5}Pr_{1/5}Sm_{1/5}O_{2-\delta}$ as a function of oxygen vacancy concentration, δ, for the fluorite (green squares) and bixbyite (purple diamonds) structures.

To explore these trends in more detail, Figure 12 presents representative projected density of states (pDOS) plots for selected oxygen vacancy concentrations in both the LN-HEO fluorite structures (top panels) and bixbyite structures (bottom panels). These snapshots illustrate how electronic structure evolves as oxygen vacancies are introduced into the lattice. These snapshots reveal the evolution of occupied and unoccupied states as a function of defect content, especially highlighting the contributions from Pr and Ce *f*-orbitals. In the lower-vacancy regime (δ ≤ 14%), the Pr 4*f*-derived states (illustrated in red) are located ~3eV deep within the valence band maximum (VBM), characteristic of the $Pr^{4+}$ oxidation state. As vacancies are introduced, particularly at 15% $V_O$ and beyond, a new Pr derived peak emerges at the top of the valence band. This spectral feature corresponds to the partially filled $4f$ orbitals of $Pr^{3+}$, indicating the onset of mixed valency. The simultaneous presence of $Pr^{3+}$ and $Pr^{4+}$ features, one near the VBM and one deeper in the valence band, reflects the internal redox behavior driven by oxygen deficiency. By 20% $V_O$, the pDOS is dominated by the $Pr^{3+}$-derived peak, confirming a near complete reduction to the trivalent state. This redistribution of electronic density results in the band gap narrowing observed in Figure 11 and is consistent with the Bader charge analysis presented in Section 3.3.1. In contrast, Ce shows greater resistance to reduction. At vacancy levels up to 20%, only a minor $Ce^{3+}$ signature appears as a faint peak near the VBM. A pronounced $Ce^{3+}$ peak, expected to manifest as a localized occupied $f$-band within the band gap, only emerges significantly at ~22% $V_O$. This delayed reduction behavior underscores Ce's relatively higher redox stability.

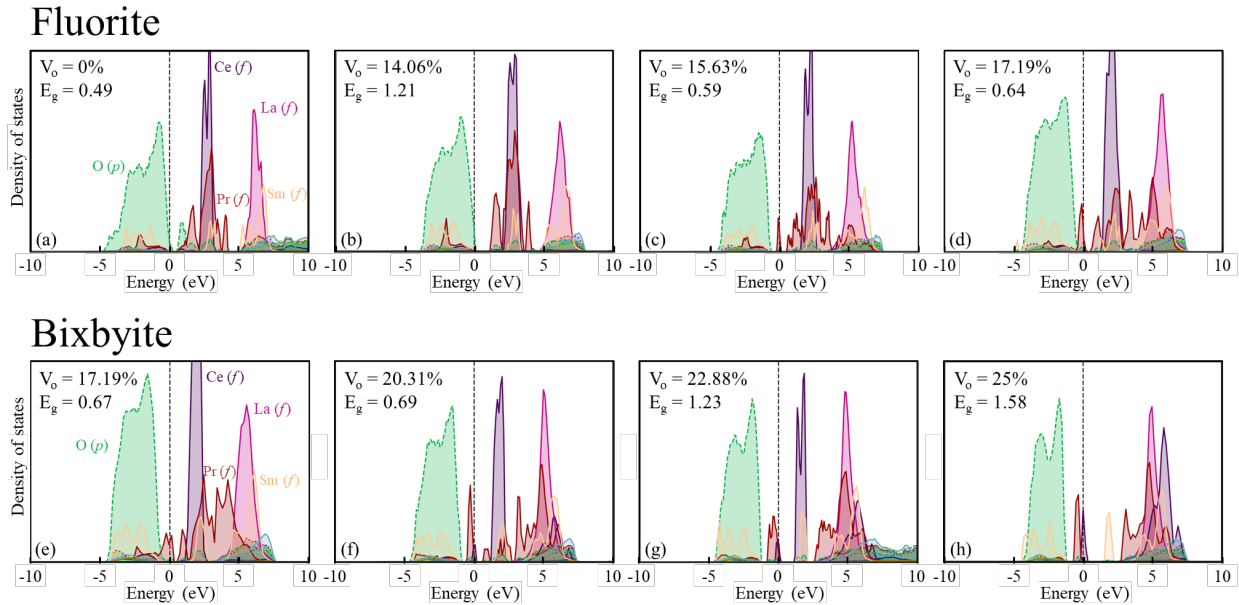

Figure 12. Projected density of states (pDOS) of the spin-up channel for $Ce_{1/5}Y_{1/5}La_{1/5}Pr_{1/5}Sm_{1/5}O_{2-\delta}$ in both (a-d) fluorite and (e-h) bixbyite structures at selected oxygen vacancy concentrations. Fermi level is set to 0 and is indicated with a black dotted line.

These trends reiterate the crucial role of internal redox in shaping the electronic landscape of LN-HEOs. The transition of Pr from a predominantly 4+ state at low δ to a mixed 3+/4+ state at intermediate δ, and eventually to fully 3+, coincides with structural destabilization, band gap modulation, and local electronic rearrangements. While Ce eventually follows a similar path, its redox shift occurs at higher vacancy levels, suggesting that Pr is the primary driver of electronic transformation within the relevant δ range (15%-22%) in agreement with the experimentally proposed Pr driven mechanism [23]. Overall, the evolution of the band structure in LN-HEOs with increasing oxygen vacancy concentration is governed by the interplay between defect chemistry and multivalent cation behavior. These DFT results reaffirm that tuning vacancy content and cation composition provides a powerful lever for engineering the electronic properties of HEOs.

## 4. CONCLUSIONS

In this work, we employed first-principles modeling to elucidate how oxygen vacancy concentration, cation composition, and internal redox jointly govern the structural, thermodynamic, and electronic responses of Ce-based lanthanide high-entropy oxides, $Ce_x(YLaPrSm)_{1-x}O_{2-\delta}$. By examining both fluorite and bixbyite phases across a wide range of Ce contents and oxygen vacancy concentrations, we reveal how the interplay between configurational disorder and vacancy chemistry dictates the stability and functional potential of LN-HEOs. As only one of these phases can potentially lead to tunable ion diffusion, it is important to understand how doping influences the structures and distortions of the system, their relative thermodynamic stabilities, and the electronic consequences of oxygen vacancy concentration and ordering.

Structurally, both fluorite and bixbyite LN-HEO phases exhibit lattice expansion with increasing oxygen deficiency, consistent with vacancy-induced local relaxation and reduction of multivalent cations with the accompanying increase in ionic radii. However, the magnitude and character of this expansion depend strongly on both phase type and cation composition. Across all compositions, the fluorite phase exhibits larger and more variable lattice expansions than bixbyite, consistent with enhanced Coulombic repulsion and heterogeneous local environments arising from randomly distributed oxygen vacancies. While a similar trend is observed in $PrO_{2-\delta}$, it is notably absent in the single-cation $CeO_{2-\delta}$ system, indicating that vacancy–vacancy interactions are amplified in multicomponent lanthanide HEOs. In these chemically disordered systems, local strain fields generated by neighboring oxygen vacancies are intensified by cation size mismatch and mixed valence, leading to greater structural variability in fluorite relative to the vacancy-ordered bixbyite phase. This contrast highlights how configurational disorder in LN-HEOs fundamentally alters vacancy-induced lattice responses compared to single-cation oxides.

In parallel, analysis of local structural distortions reveals heightened polyhedral asymmetry, particularly in the disordered fluorite phase, underscoring how vacancy distribution disrupts local coordination environments. Local bonding analysis further confirms that vacancy disorder drives substantial broadening of LN–O bond distributions and pronounced polyhedral distortions, while vacancy ordering in bixbyite suppresses these effects only at high vacancy concentrations. Among the constituent cations, Pr exhibits the strongest vacancy-coupled redox response, with the onset of $Pr^{4+} \rightarrow Pr^{3+}$ reduction coinciding with abrupt bond elongation and increased polyhedral asymmetry, whereas La, Sm, and Y maintain stable trivalent states across all $\delta$.

Electronic structure and Bader charge analyses indicated: La, Sm, and Y remain stable in the trivalent state across the studied vacancy range, Ce exhibits a suppressed reduction behavior by retaining a dominant $Ce^{4+}$ character up to ~22% $V_O$, and Pr undergoes a pronounced transition to a mixed-valence 3+/4+ state beginning near 14% $V_O$. These redox dynamics are closely mirrored in the electronic structure. Correspondingly, pDOS reveal progressive band gap narrowing primarily by the emergence of $Pr^{3+}$-derived states near the valence band maximum. Whereas, the $Ce^{3+}$ signatures emerge only at higher vacancy levels, reinforcing its more stable 4+ oxidation state in the high-entropy environment.

Thermodynamic analysis reveals that both equimolar fluorite and bixbyite LN-HEOs exhibit a second-order, nonlinear response of formation energy on oxygen vacancy concentration, with minima occurring near 20-25% $V_O$, indicating an intrinsic energetic preference for high vacancy concentrations in both structures, independent of phase. Across Ce compositions, the 25% $V_O$ formation enthalpy of fluorite forms a nearly flat, Ce-independent curve that closely parallels the bixbyite line but remains consistently ~0.1 eV/atom higher. Since the two structures differ only in the spatial arrangement of oxygen vacancies, this offset directly quantifies the enthalpic penalty associated with random vacancy placement and demonstrates an inherent energetic preference for vacancy ordering in the bixbyite phase at high defect levels.

Despite this enthalpic bias, convex-hull and free-energy analyses show that, while bixbyite is enthalpically favored across all Ce concentrations, incorporating the configurational entropy associated with disordered oxygen vacancies substantially lowers the free energy of fluorite, stabilizing the disordered phase above approximately 30% Ce. This entropy-driven stabilization explains the experimentally observed sensitivity of phase formation to composition and synthesis conditions and highlights the delicate thermodynamic balance between disordered fluorite and partially ordered bixbyite structures in LN-HEOs. Cooperatively, this demonstrates that phase stability in cerium-based LN-HEOs is governed by a competition between vacancy-ordering enthalpy and anion configurational entropy, underscoring the critical role of temperature and compositional tuning in designing vacancy-tolerant, fluorite-based oxide electrolytes.

Collectively, these findings provide a mechanistic framework for understanding how cation disorder, oxygen vacancy chemistry, and configurational entropy interact to stabilize competing phases in lanthanide HEOs. The decoupling of redox behavior from structural phase stability, the clear enthalpic preference for ordered vacancies, and the strong entropy-driven stabilization of fluorite all highlight the tunability of these materials through composition and synthesis conditions. By clarifying how vacancy distribution governs both structure and electronic response, this work offers guiding principles for rationally engineering LN-HEOs for functional applications such as solid electrolytes, memristive devices, and redox-active catalysts. The insights developed here emphasize the critical role of anion-sublattice disorder in designing next-generation high-entropy ceramics with tailored transport and electronic properties.

**Acknowledgements:**
The authors acknowledge the use of facilities and instrumentation supported by NSF through the Pennsylvania State University Materials Research Science and Engineering Center [DMR-2011839]. Computations for this research were performed on the Pennsylvania State University's Institute for Computational and Data Sciences' Roar supercomputer.

**CRediT authorship contribution statement**
**Mary K. Caucci:** Writing – original draft, Writing – review & editing, Visualization, Validation, Methodology, Investigation, Formal analysis, Data curation, Conceptualization. **Billy Yang:** Writing – review & editing, Conceptualization. **Gerald Bejger:** Writing – review & editing, Conceptualization. **Jacob Sivak:** Writing – review & editing, Conceptualization, Validation. **Saeed Almishal:** Writing – review & editing, Conceptualization. **Christina Rost:** Writing – review & editing, Funding acquisition, Conceptualization. **Jon-Paul Maria:** Writing – review & editing, Funding acquisition, Conceptualization. **Susan Sinnott:** Writing – review & editing, Supervision, Resources, Funding acquisition, Conceptualization.

# SUPPLEMENTAL MATERIAL

## Section S.1: Tables

Table S.1. $Ce_x(YLaPrSm)_{1-x}O_{2-\delta}$ compositions and their respective supercells used in density functional theory (DFT) simulations, where x is the Ce concentration in mol, z is the number of oxygens removed from the DFT supercell, and $\delta$ is the oxygen non-stoichiometry.

| Compositions | Oxygen non-stoichiometry, $\delta$ | Supercell formula | $\delta$ of supercell with lowest $\Delta H_f$ | |
|---|---|---|---|---|
| | | | Fluorite | Bixbyite |
| $Ce_{0.22}Y_{0.19}La_{0.19}Pr_{0.22}Sm_{0.19}O_{2-\delta}$ | 0.0, 0.09, 0.19, 0.25, 0.28, 0.31, 0.34, 0.41, 0.43, 0.5, 0.59 | $Ce_7Y_6La_6Pr_7Sm_6O_{64-z}$ | 0.34 | 0.5 |
| $Ce_{0.25}Y_{0.19}La_{0.19}Pr_{0.19}Sm_{0.19}O_{2-\delta}$ | 0.0, 0.28, 0.31, 0.34, 0.41, 0.5 | $Ce_8Y_6La_6Pr_6Sm_6O_{64-z}$ | 0.34 | 0.5 |
| $Ce_{0.31}Y_{0.19}La_{0.16}Pr_{0.19}Sm_{0.16}O_{2-\delta}$ | 0.0, 0.25, 0.28, 0.31, 0.34, 0.41, 0.5 | $Ce_{10}Y_6La_5Pr_6Sm_5O_{64-z}$ | 0.41 | 0.43 |
| $Ce_{0.34}Y_{0.16}La_{0.16}Pr_{0.19}Sm_{0.16}O_{2-\delta}$ | 0.0, 0.22, 0.25, 0.28, 0.31, 0.34, 0.41, 0.43, 0.5 | $Ce_{11}Y_5La_5Pr_6Sm_5O_{64-z}$ | 0.41 | 0.43 |
| $Ce_{0.38}Y_{0.16}La_{0.16}Pr_{0.16}Sm_{0.16}O_{2-\delta}$ | 0.0, 0.22, 0.25, 0.28, 0.31, 0.34, 0.41, 0.43, 0.5 | $Ce_{12}Y_5La_5Pr_5Sm_5O_{64-z}$ | 0.25 | 0.41 |
| $Ce_{0.50}Y_{0.13}La_{0.13}Pr_{0.13}Sm_{0.13}O_{2-\delta}$ | 0.0, 0.19, 0.22, 0.25, 0.28, 0.41, 0.5 | $Ce_{16}Y_4La_4Pr_4Sm_4O_{64-z}$ | 0.25 | 0.25 |
| $Ce_{0.81}Y_{0.03}La_{0.03}Pr_{0.09}Sm_{0.03}O_{2-\delta}$ | 0.0, 0.06, 0.19, 0.25, 0.41, 0.5 | $Ce_{26}Y_1La_1Pr_3Sm_1O_{64-z}$ | 0.19 | 0.25 |
| $Ce_{0.88}Y_{0.03}La_{0.03}Pr_{0.03}Sm_{0.03}O_{2-\delta}$ | 0.0, 0.06, 0.19, 0.25, 0.41, 0.5 | $Ce_{28}Y_1La_1Pr_1Sm_1O_{64-z}$ | 0.09 | 0.31 |
| $CeO_{2-\delta}$ | 0.0, 0.09, 0.19, 0.31, 0.41, 0.5, 0.59 | $Ce_{32}O_{64-z}$ | 0.34 | 0.5 |
| $PrO_{2-\delta}$ | 0.0, 0.09, 0.19, 0.31, 0.41, 0.5, 0.59 | $Pr_{32}O_{64-z}$ | | |

Table S.2. Solid elemental reference states.

| Symbol | Spacegroup | Structure |
|---|---|---|
| Y | $P6_3/mmc$ | hcp |
| La | $P6_3/mmc$ | dhcp |
| Ce | $Fm\bar{3}m$ | fcc |
| Pr | $P6_3/mmc$ | dhcp |
| Sm | $R\bar{3}m$ | rhomb, αSm-type |

## Section S.2: Structural properties

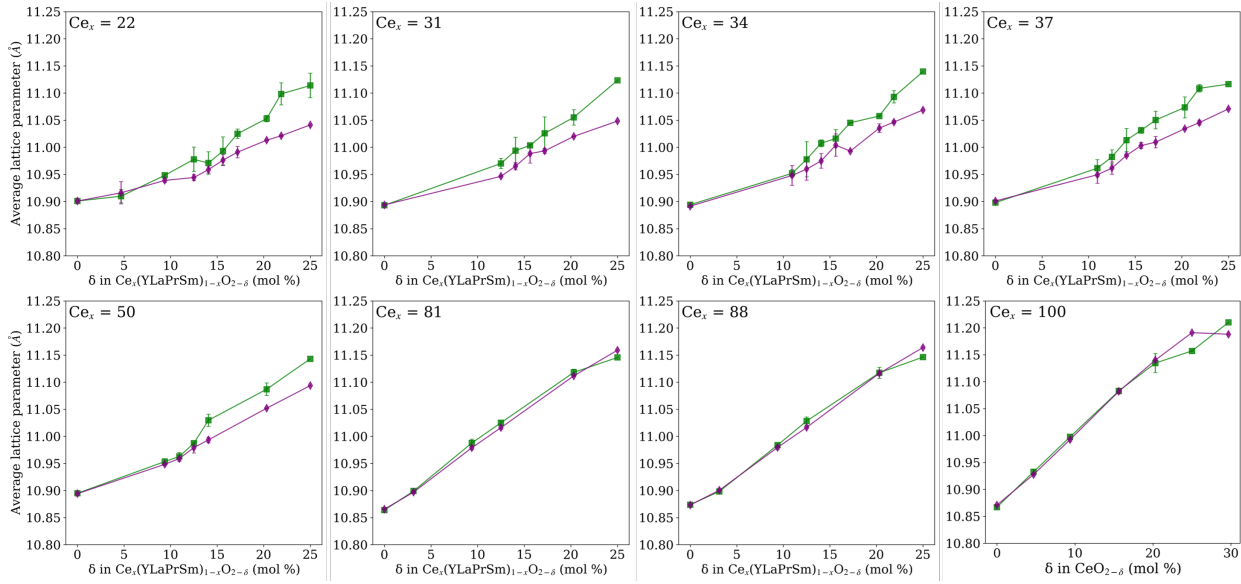

Figure S.1. Computed supercell lattice constants of fluorite (green squares) and bixbyite (purple diamonds) LN-HEO phases as a function of oxygen vacancy concentration. Error bars are the standard deviation between calculations.

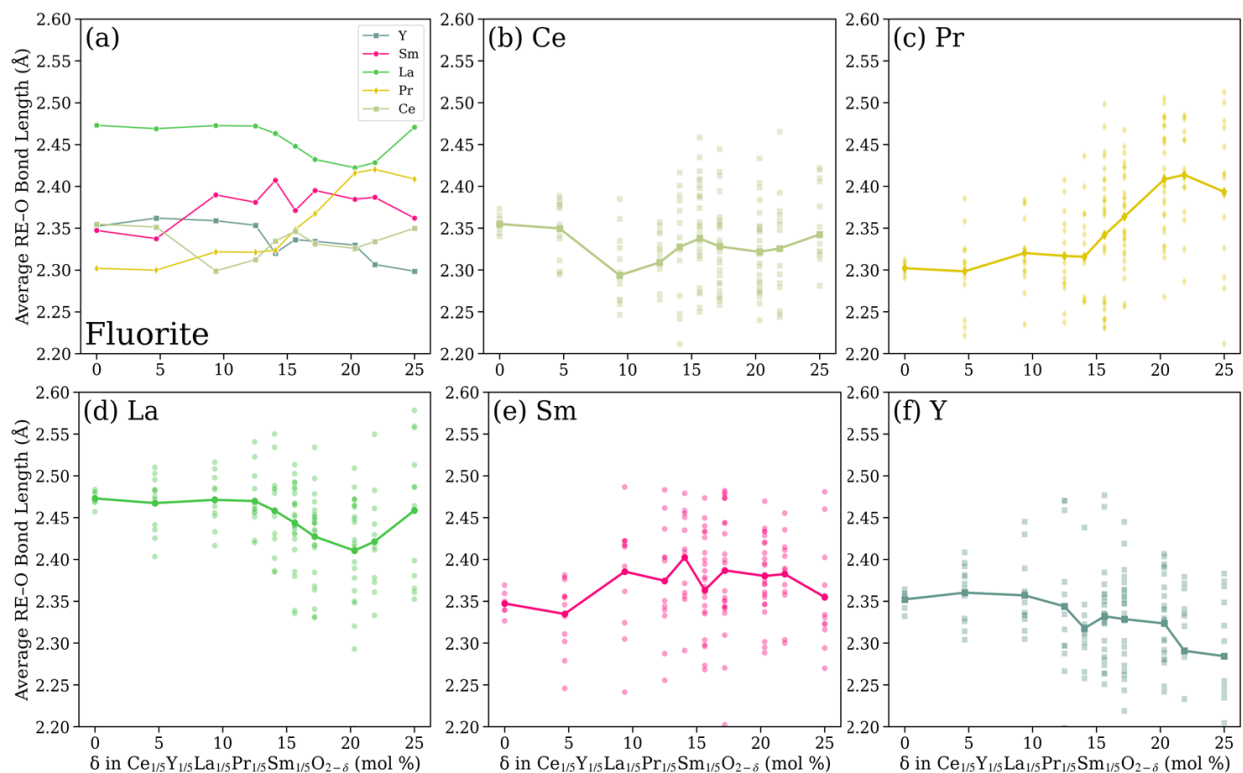

Figure S.2. (a) Average LN-O bond lengths (in Å) for the equimolar fluorite structure as a function of oxygen vacancy concentration. (b-f) Individual LN-O bond lengths for Ce, Pr, La, Sm, and Y, with solid lines connecting average values to guide the eye.

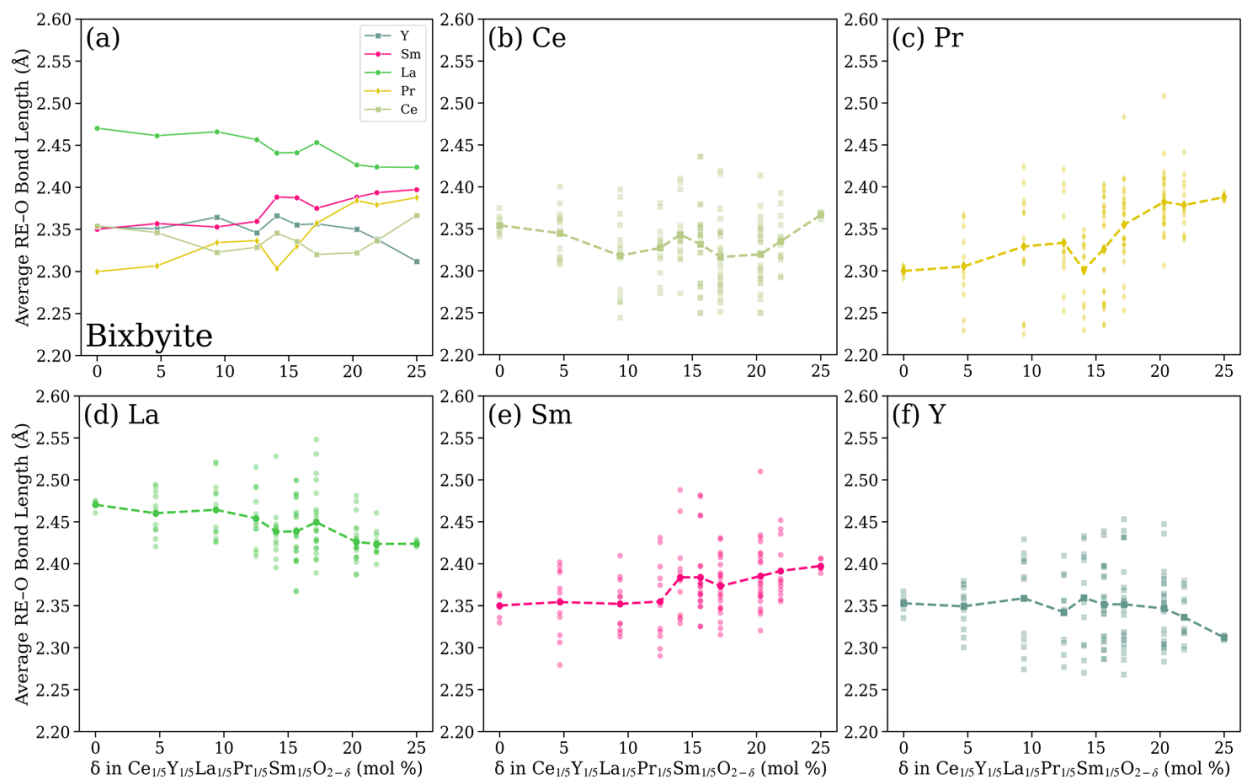

Figure S.3. (a) Average LN-O bond lengths (in Å) for the equimolar bixbyite structure as a function of oxygen vacancy concentration. (b-f) Individual LN-O bond lengths for Ce, Pr, La, Sm, and Y, with solid lines connecting average vales to guide the eye.

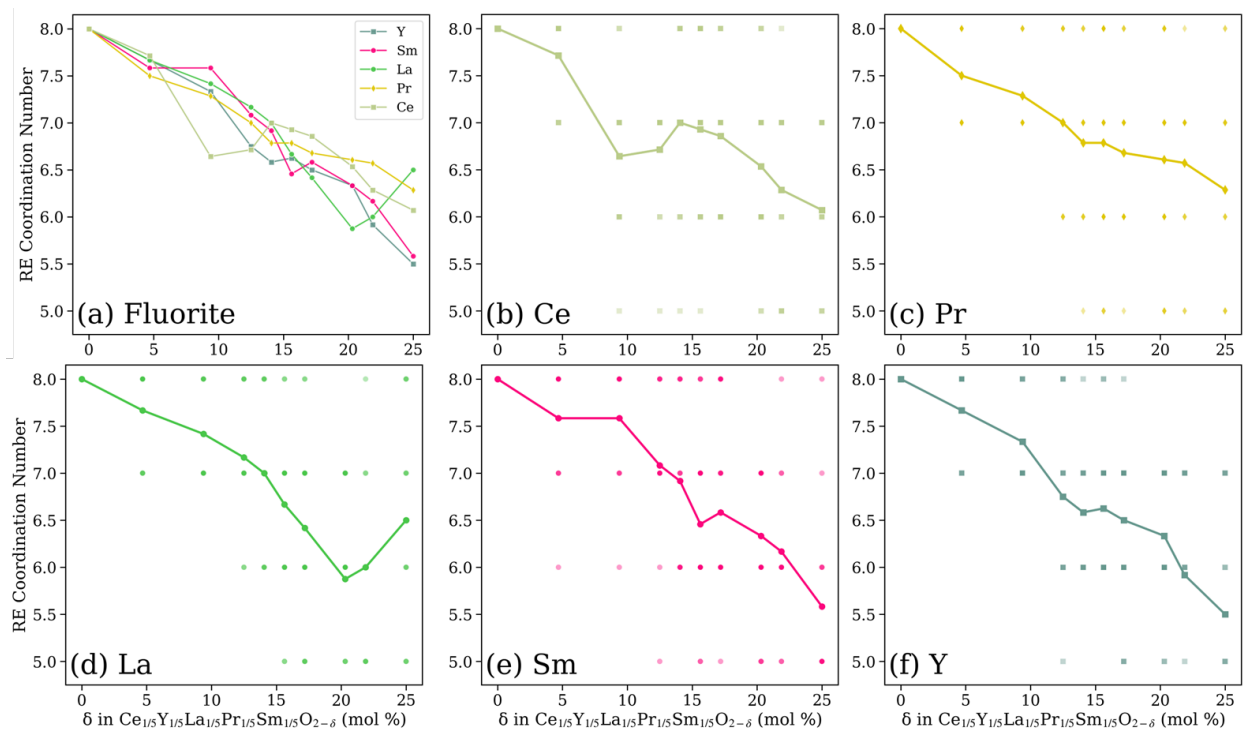

Figure S.4. (a) Average LN-O coordination number for the equimolar fluorite structure as a function of oxygen vacancy concentration. (b-f) Coordination number distributions for Ce, Pr, La, Sm, and Y with the lines connecting the average value. The decline in coordination number with increasing $\delta$ indicates the progressive removal of oxygen atoms from the local environment of LN cations in the disordered structure.

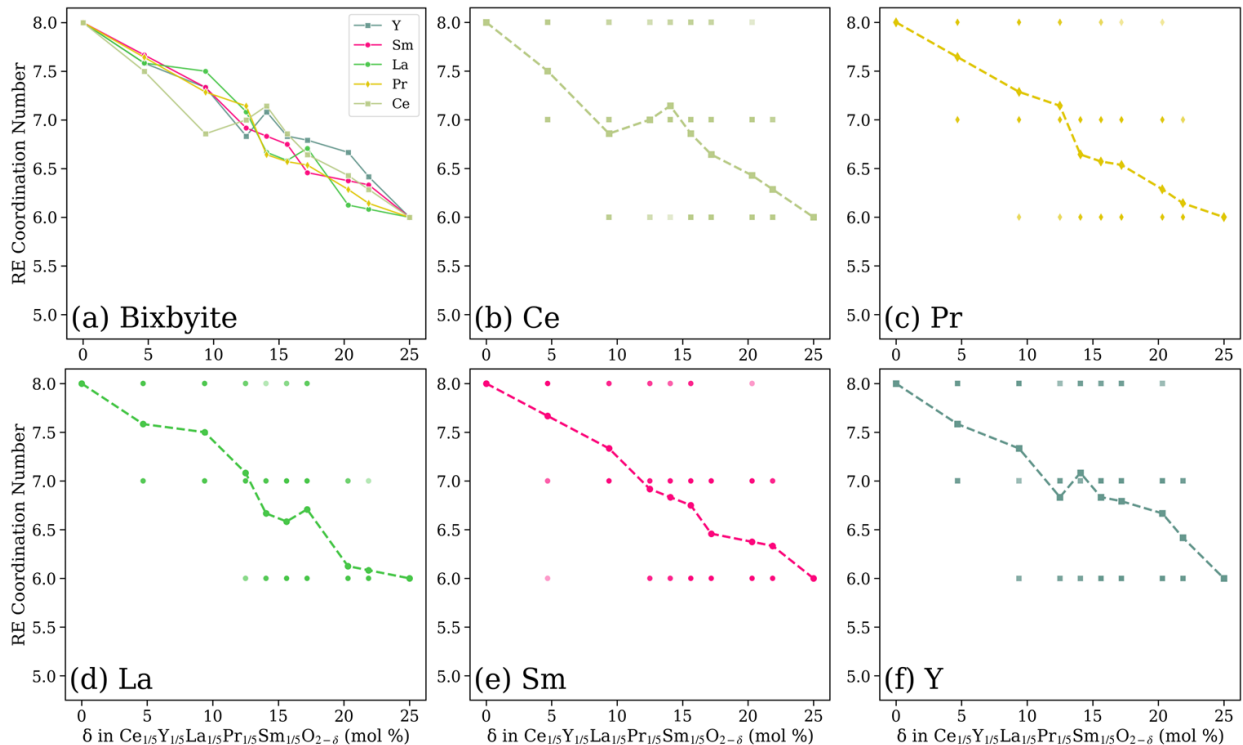

Figure S.5. (a) Average LN-O coordination number for the equimolar bixbyite structure across oxygen vacancy concentrations. (b-f) Coordination number distributions for Ce, Pr, La, Sm, and Y with the lines connecting the average value. Bixbyite exhibits less variation due to its more uniform and ordered vacancy sublattice, which limits the degree of local coordination asymmetry compared to fluorite.

# Fluorite

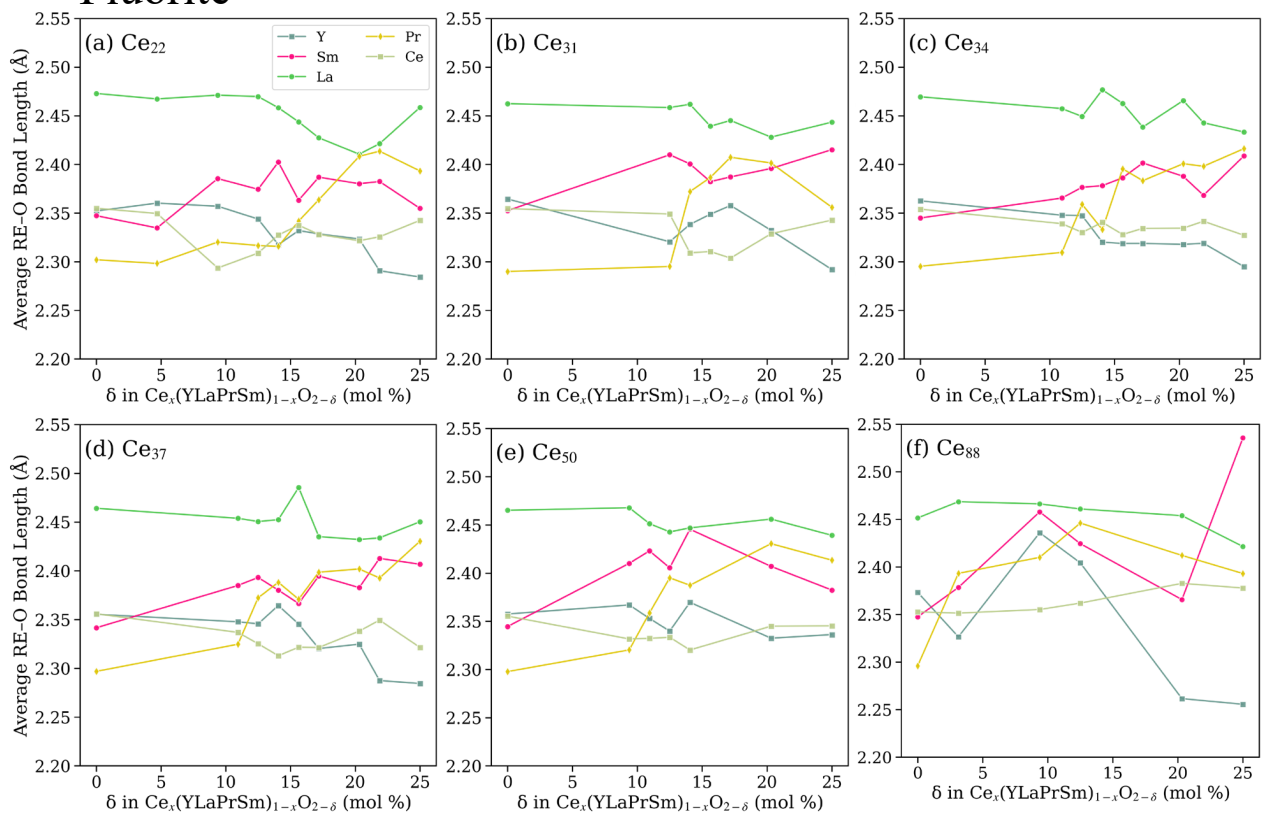

Figure S.6. Average LN–O bond lengths (in Å) for fluorite $Ce_x(YLaPrSm)_{1-x}O_{2-\delta}$ structures as a function of oxygen non-stoichiometry ($\delta$). Each point represents the mean bond length across multiple supercells.

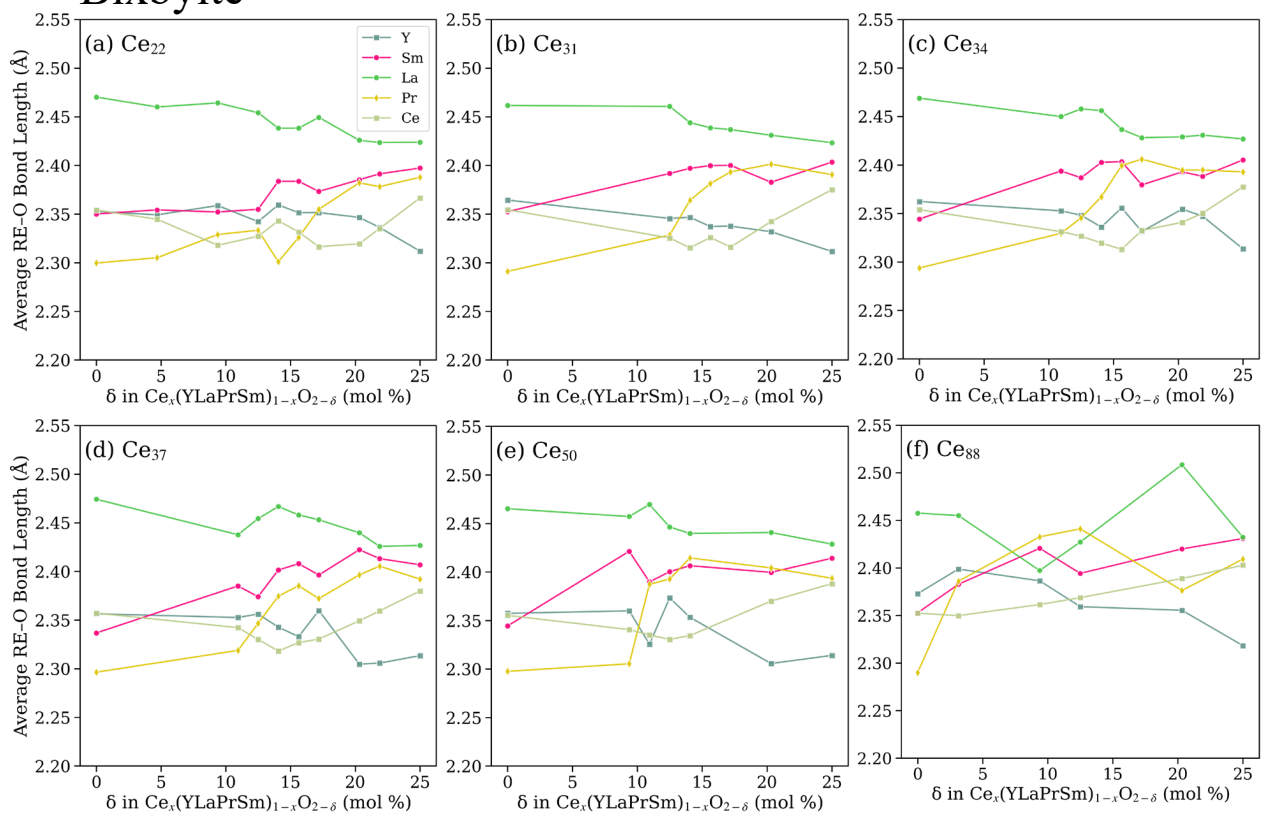

Figure S.7. Average LN–O bond lengths (in Å) for bixbyite $Ce_x(YLaPrSm)_{1-x}O_{2-\delta}$ structures as a function of oxygen non-stoichiometry ($\delta$). Each point represents the mean bond length across multiple supercells.

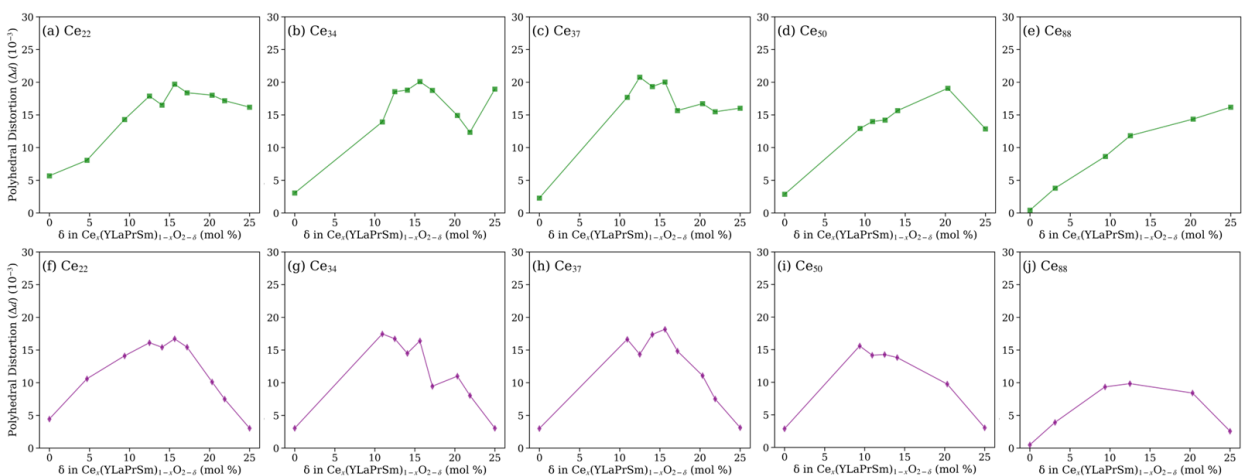

Figure S.8. Average polyhedral distortion index ($\Delta d$) for LN-centered coordination environments in (a, c) fluorite and (b, d) bixbyite structures as a function of oxygen vacancy concentration.

**Section S.3: Thermodynamics**

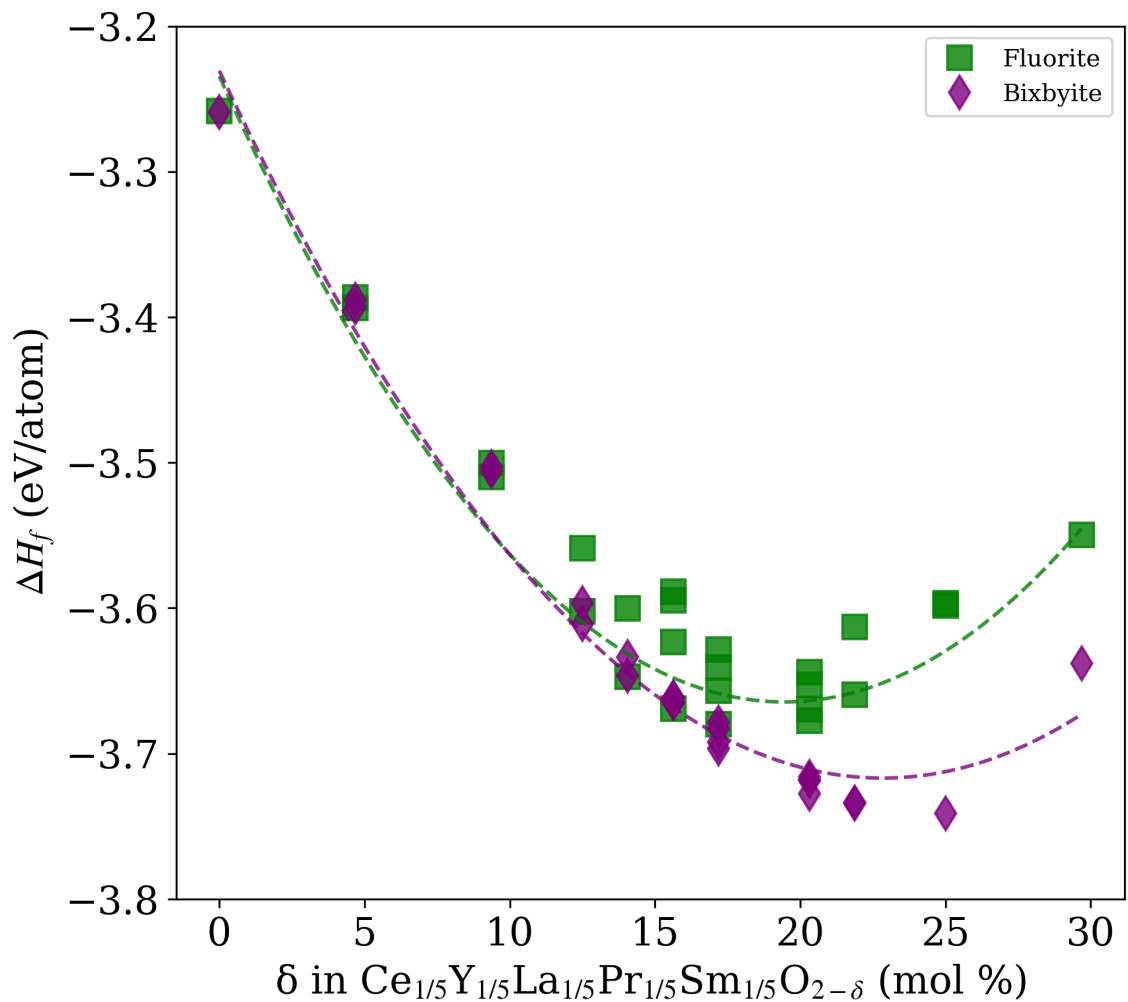

Figure S.9. Formation enthalpy $\Delta H_f$ per atom as a function of oxygen vacancy (non-stoichiometry, $\delta$) for $Ce_{1/5}Y_{1/5}La_{1/5}Pr_{1/5}Sm_{1/5}O_{2-\delta}$. Green squares (fluorite) and purple diamonds (bixbyite) show DFT calculated formation enthalpies across multiple oxygen vacancy configurations. Dotted curves represent second-order polynomial least-squares fits to the minimum energies of each phase, used to estimate the vacancy concentration at which each structure reaches its thermodynamic minimum.

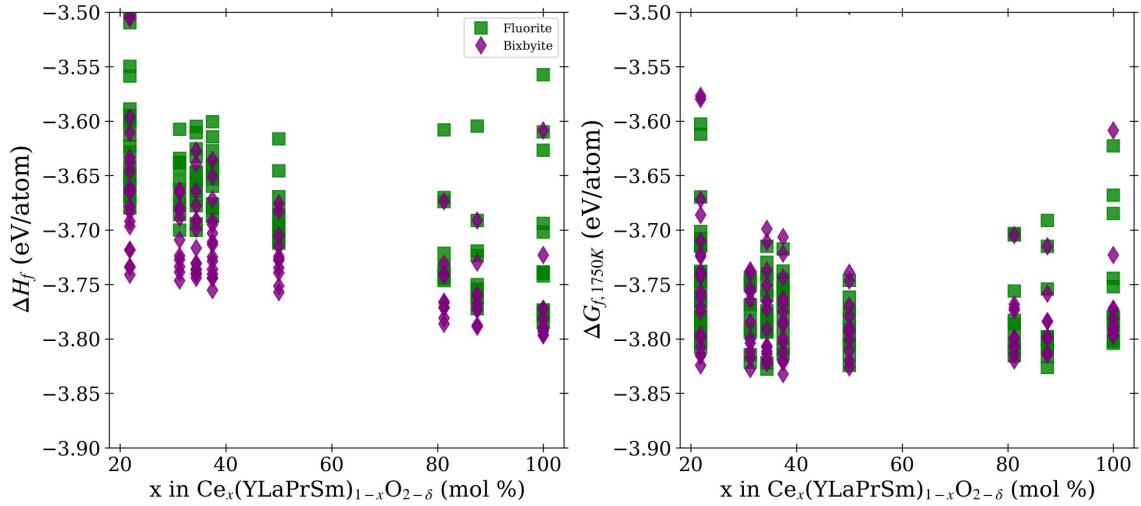

Figure S.10. (right) Formation enthalpy $\Delta H_f$ per atom and (left) free energy $\Delta G_{f,1750K}$ as a function of Ce content for fluorite (green squares) and bixbyite (purple diamonds) $Ce_x(YLaPrSm)_{1-x}O_{2-\delta}$ structures with considered oxygen non-stoichiometries $\delta$.

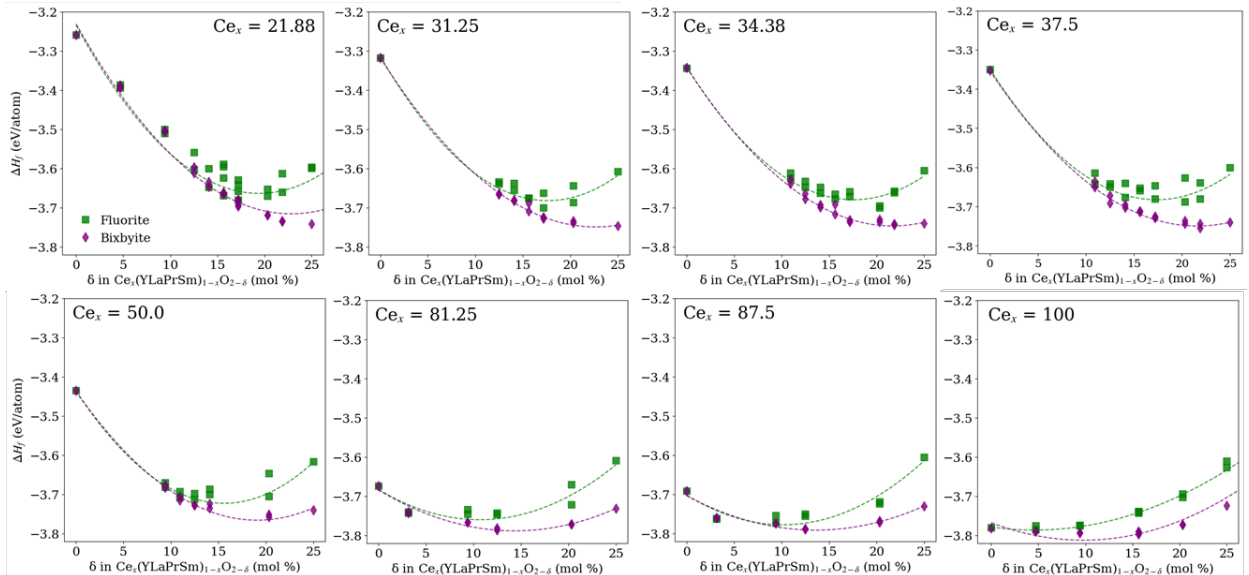

Figure S.11. Formation enthalpy $\Delta H_f$ per atom as a function of $V_O$ content for fluorite (green squares) and bixbyite (purple diamonds) $Ce_x(YLaPrSm)_{1-x}O_{2-\delta}$. Discrete points represent DFT-computed values, while dotted lines indicate second-order polynomial least-squares fits to $\Delta H_f$ minimum used to extract the vacancy concentration corresponding to the minimum enthalpy at each Ce content.

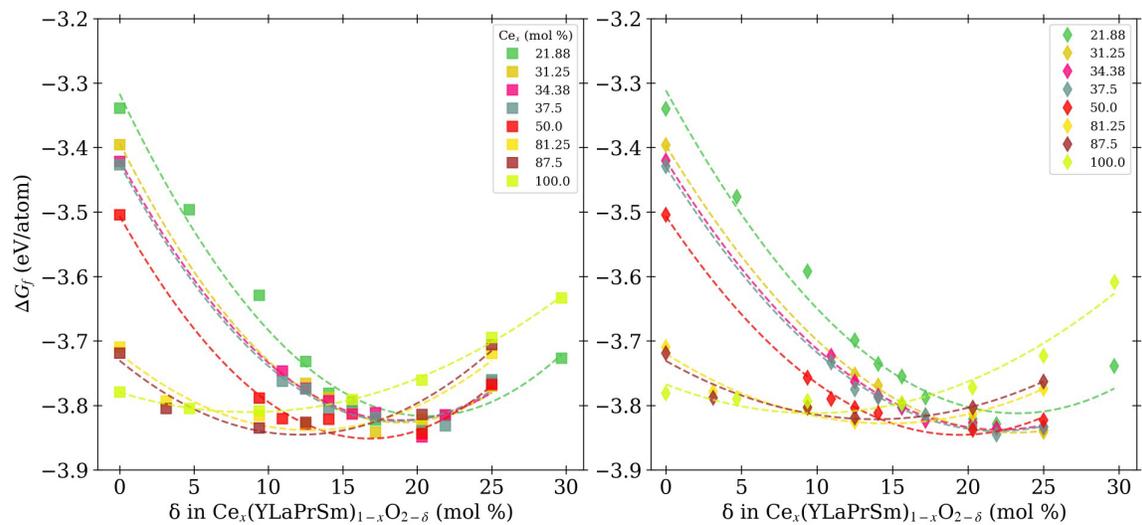

Figure S.12. Free energy $\Delta G_f$ per atom as a function of oxygen vacancy (non-stoichiometry, $\delta$) at T = 1750K for (a) fluorite and (b) bixbyite $Ce_x(YLaPrSm)_{1-x}O_{2-\delta}$ structures at the considered Ce concentrations. Discrete points represent DFT-computed energies, while dotted lines indicate second-order polynomial least-squares fits used to extract the vacancy concentration corresponding to the minimum enthalpy at each Ce content.

**Section S.3: Electronic Structure**

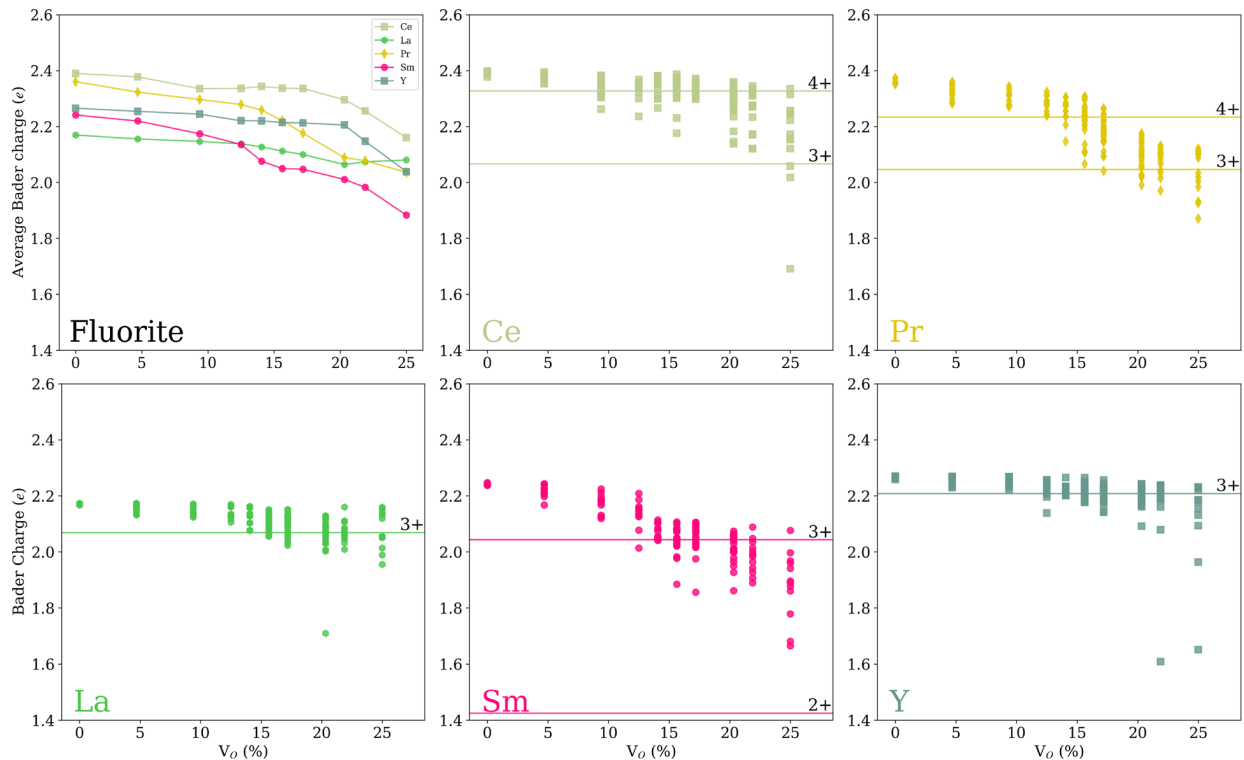

Figure S.13. (a) Average and (b-f) individual Bader charges of cations in $Ce_{1/5}Y_{1/5}La_{1/5}Pr_{1/5}Sm_{1/5}O_{2-\delta}$ in the fluorite phase. Horizontal solid lines mark reference Bader charges from the corresponding binary oxides: +3 for La, Sm, Y and both +3/+4 for Ce and Pr.

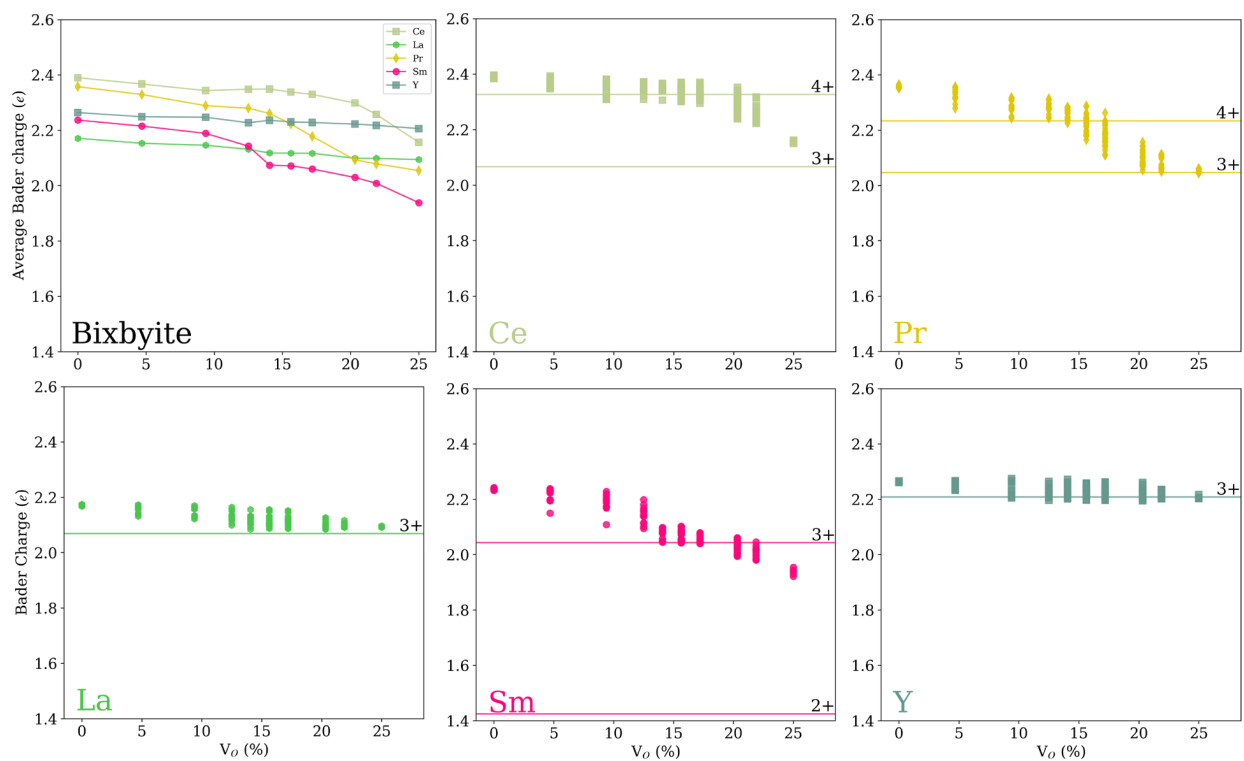

Figure S.14. (a) Average and (b-f) individual Bader charges of cations in $Ce_{1/5}Y_{1/5}La_{1/5}Pr_{1/5}Sm_{1/5}O_{2-\delta}$ in the bixbyite phase. Horizontal solid lines mark reference Bader charges from the corresponding binary oxides: +3 for La, Sm, Y and both +3/+4 for Ce and Pr.

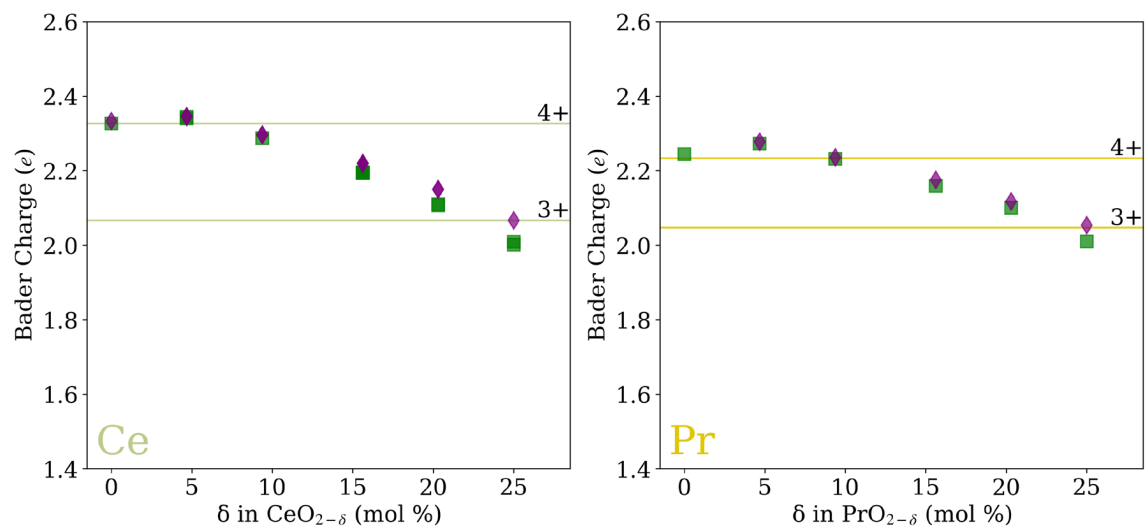

Figure S.15. Average Bader charges of cations in (right) $CeO_{2-\delta}$ and (left) $PrO_{2-\delta}$ across increasing oxygen vacancies for fluorite (green squares) and bixbyite (purple diamonds) structures. Horizontal solid lines mark reference Bader charges from the corresponding binary oxide

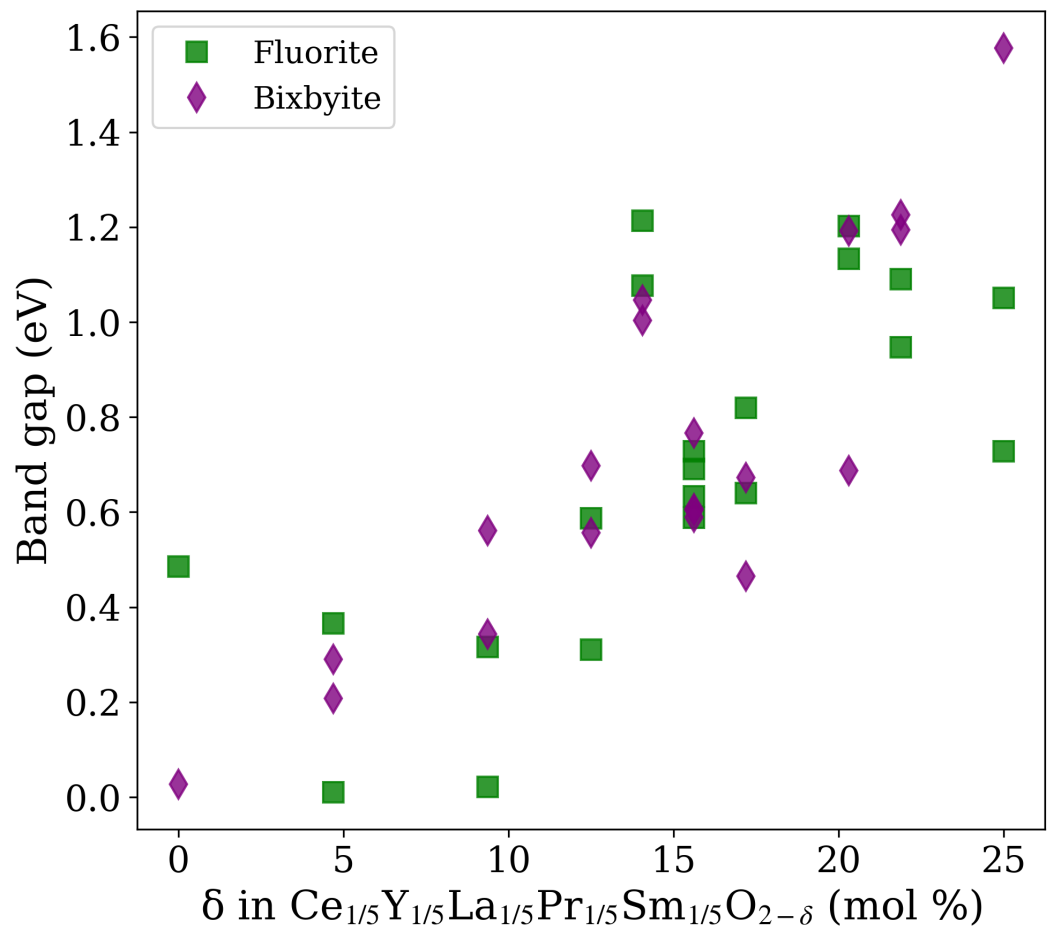

Figure S.16. DFT predicted band gap (in eV) of $Ce_{1/5}Y_{1/5}La_{1/5}Pr_{1/5}Sm_{1/5}O_{2-\delta}$ as a function of oxygen vacancy concentration, $\delta$, for the fluorite (green squares) and bixbyite (purple diamonds) structures.